\documentclass{elsart}

\usepackage{natbib}
\usepackage{epsfig}
\usepackage{amsmath}
\usepackage{amssymb}
\usepackage{graphicx}
\usepackage{subfigure}
\newcommand \be{\begin{equation}}
\newcommand \ba{\begin{eqnarray}}
\newcommand \ee{\end{equation}}
\newcommand \ea{\end{eqnarray}}

\begin{document}
\begin{frontmatter}

\title{Importance of Positive Feedbacks and Over-confidence
in a Self-Fulfilling Ising Model of Financial Markets}

\author[ucla,nice]{\small{Didier Sornette}\thanksref{EM}},
\author[ecust]{\small{Wei-Xing Zhou}}

\address[ucla]{Institute of Geophysics and Planetary Physics and
Department of Earth and Space Sciences, University of California,
Los Angeles, CA 90095}

\address[nice]{Laboratoire de Physique de la Mati\`ere Condens\'ee, CNRS
UMR 6622 and Universit\'e de Nice-Sophia Antipolis, 06108 Nice
Cedex 2, France}

\address[ecust]{State Key Laboratory of Chemical Reaction Engineering,\\
East China University of Science and Technology, Shanghai 200237,
China}

\thanks[EM]{Corresponding author. Department of Earth and Space
Sciences and Institute of Geophysics and Planetary Physics,
University of California, Los Angeles, CA 90095-1567, USA. Tel:
+1-310-825-2863; Fax: +1-310-206-3051. {\it E-mail address:}\/
sornette@moho.ess.ucla.edu (D. Sornette)\\
http://www.ess.ucla.edu/faculty/sornette/}

\begin{abstract}

Following a long tradition of physicists who have noticed that the
Ising model provides a general background to build realistic
models of social interactions, we study a model of financial price
dynamics resulting from the collective aggregate decisions of
agents. This model incorporates imitation, the impact of external
news and private information. It has the structure of a dynamical
Ising model in which agents have two opinions (buy or sell) with
coupling coefficients which evolve in time with a memory of how
past news have explained realized market returns. We study two
versions of the model, which differ on how the agents interpret
the predictive power of news. We show that the stylized facts of
financial markets are reproduced only when agents are
over-confident and mis-attribute the success of news to predict
return to herding effects, thereby providing positive feedbacks
leading to the model functioning close to the critical point. Our
model exhibits a rich multifractal structure characterized by a
continuous spectrum of exponents of the power law relaxation of
endogenous bursts of volatility, in good agreement with previous
analytical predictions obtained with the multifractal random walk
model and with empirical facts.

Dynamic Games; Evolutionary Games

\end{abstract}

\end{frontmatter}

\newpage

Biographies:

Didier SORNETTE is a professor at the University of
California, Los Angeles and a research director of CNRS, the
French National Center for Scientific Research. He received his
PhD in Statistical Physics from the University of Nice, France.
His current research focuses on the modelling and prediction of
catastrophic events in complex systems, with applications to
finance, economics, seismology, geophysics, and biology.

Wei-Xing ZHOU is an associate research professor at East China
University of Science and Technology. He received his PhD in
Chemical Engineering from East China University of Science and
Technology in 2001. His current research interest focuses on the
modelling and prediction of catastrophic events in complex
systems.

{\large{\bf{Acknowledgements}}}

We are grateful to Carlos Pedro Gon\c{c}alves,
Jean-Fran\c{c}ois Muzy and Matthieu Wyart for stimulating
discussions. The research by W.-X. Zhou was supported by
NSFC/PetroChina jointly through a major project on multiscale
methodology (No. 20490200).

\newpage

\section{Introduction}
\label{s1:introduction}

Many works borrow concepts from the theory of the Ising models and
of phase transitions to model social interactions and organization
\citep[see,~e.g.,][]{PhysicsToday,Montroll}. In particular,
\citet{Orlean1,Orlean2,Orlean3,Orlean4,Orlean5,Orlean6} has
captured the paradox of combining rational and imitative behavior
under the name ``mimetic rationality,'' by developing models of
mimetic contagion of investors in the stock markets which are
based on irreversible processes of opinion forming. As recalled in
the Appendix, the dynamical updating rules of the Ising model are
obtained in a natural way to describe the formation of decisions
of boundedly rational agents \citep{RS00}. The Ising  model is one
of the simplest models describing the competition between the
ordering force of imitation or contagion and the disordering
impact of private information or idiosyncratic noise
\citep{Ising1}. In the same class of minimal models of emergent
social behaviors is the model of herding based on percolation
clusters proposed by \citet{Contboudaaq}.

Starting with a framework suggested by
\citet{Blume1,Blume2,BrockHomme1,Durlauf1,Durlauf2,Durlauf3,Durlauf4,Durlauf5,Durlauf6},
\citet{Phanetal} summarize the formalism starting
with different implementation of the agents' decision processes whose
aggregation is inspired from statistical mechanics to
account for social influence in individual decisions.
\citet{LuxMarchesi,LuxMarchesi2,BrockHomme4,KirmTey} have also
developed related models in which agents' successful forecasts
reinforce the forecasts. Such models have been found to generate swings
in opinions, regime changes and long memory. An essential feature of
these models is that agents are wrong for some of the time, but,
whenever they are in the majority they are essentially right. Thus they
are not systematically irrational \citep{KirmRev}.

Here, we study a model of interacting agents buying and selling a
single financial asset based on an extension of the Ising model.
The agents make their decision based upon the combination of three
different information channels: mutual influences or imitation,
external news and idiosyncratic judgements. Agents update their
willingness to extract information from the other agents' behavior
based on their assessment of how past news have explained market
returns. Agents update their propensity to herding according to
what degree the news have been successful in predicting returns.
We  distinguish between two possible updating rules: rational and
irrational. In the rational version, agents decrease their
propensity to imitate if news have been good predictors of returns
in the recent past. In the irrational version, agents
mis-attribute the recent predictive power of news to their
collective action, leading to positive self-reinforcement of
imitation. We show that the model can reproduce the major
empirical stylized facts of financial stock markets only when the
updating of the strength of imitation is irrational, providing a
direct test and the evidence for the importance of misjudgement of
agents biased toward herding.

Section 2 specifies the model and compares it with previous
related versions. Section 3 presents the results of exhaustive
searches in the space of the major parameters of the two versions
of the model. We describe in turn the distributions of returns at
multiple time scales, the auto-correlations of the returns and of
the volatility (absolute value of the returns) at different time
scales, the multifractal properties of the structure functions of
the absolute values of returns and their consequences in the
characteristic relaxation of the volatility after bursts of
endogenous versus exogenous origins. Section 4 concludes.

\section{Model of imitation versus news impact}

\subsection{Definition of the model \label{jfjs}}

We study the following model of $N$ agents interacting within a
network ${\mathcal{N}}$ (taken here for simplicity as the set of
nodes linked by nearest-neighbor bonds on the square lattice; this
implies that an agent sitting at a node interacts directly only
with her four neighbors). At each time step $t$, each agent $i$
places a buy ($s_i(t)=+1$) or sell ($s_i(t)=-1$) order. Her
decision $s_i(t)$ is determined by the following process
\begin{equation}
 s_i(t)={\rm{sign}}\left[\sum_{j\in{\mathcal{N}}}K_{ij}(t) {\rm E}[s_j](t)+
 \sigma_i(t) G(t) +\epsilon_i(t)\right]~,
 \label{Eq:Sit22}
\end{equation}
where ${\rm E}[s_j](t)$ is the expectation formed by agent $i$ on
what will be the decision of agent $j$ at the same time $t$. The
left-hand-side and the right-hand-side are in principle evaluated
simultaneously at the same time to capture the anticipation by a
given agent $i$ of the actions of the other agents which are going
to determine the change of the market price from $t-1$ to $t$.
Indeed, we assume that the decisions $s_i(t)$ are formed slightly
before $t$, in the period from $t-1$ to $t$, when the news $G(t)$
has become available in the interval from $t-1$ to $t$, and are
then converted into price at $t$ by the market clearing process.
In principle, the best strategy for agent $i$ is indeed to base
her action for the next investment period on her best guess of the
{\it present} action of all other agents for the next investment
period (see below).

Expression (\ref{Eq:Sit22}) embodies three contributions to the
decision making process of agent $i$:
\begin{itemize}
\item {\bf Imitation through the term
$\sum_{j\in{\mathcal{N}}}K_{ij}(t) {\rm E}[s_j](t)$}. The kernel
$K_{ij}$ is the relative propensity of the trader $i$ to be
contaminated by the sentiment of her friend $j$ (coefficient of
influence of $j$ on $i$). In other words, $K_{ij}(t)$ quantifies
the strength of the influence of agent $j$'s expected decision on
the decision of agent $i$, which evolves with time as we soon
specify. Due to the heterogeneity of the traders, $K_{ij}\ne
K_{ji}$, generally. The sum $\sum_{j\in{\mathcal{N}}}$ is carried
over all agents $j$ who are in direct contact with agent $i$.

\item {\bf Impact of external news through the term $\sigma_i(t)
G(t)$}. $G(t)$ quantifies the impact of the external news $I(t)$
on the decision of agent $i$. We follow the specification of the
artificial stock market model formulated by \citet{G03AFM}, and
assume that $I(t)$ follows a standard normal distribution and
\begin{equation}
 G(t)=\left\{
 \begin{array}{rl}
 1, & {\rm{if}}~ I(t)>0~,\\
 -1, & {\rm{if}}~ I(t) \leq 0~.
 \end{array}
 \right.
\end{equation}
$\sigma_i(t)$ quantifies the relative impact at time $t$ of the
news' positive or negative outlook on the decision process of
agent $i$. In other words, $\sigma_i$ is the relative sensitivity
of agent's sentiment to the news.

\item {\bf Idiosyncratic judgement associated with private
information}. $\epsilon_i(t)$ embodies the idiosyncratic content
of the decision of agent $i$ accounting for the interpretation of
her own private information. We take it as being normally
distributed around zero with a constant (homogeneous) standard
deviation equal to unity, without loss of generality, since the
relative strength of the three terms is already captures by the
units of $K_{ij}$'s and $\sigma_i$'s.
\end{itemize}

The market price is updated according to \be p(t)=p(t-1)
\exp[r(t)]~, \label{ngmldl} \ee so that $p(t)$ and $r(t)$ are
known at the end of the interval from $t-1$ to $t$. The return
$r(t)$ is determined according to
\begin{equation}
 r(t) = \frac{\sum_{i\in{\mathcal{N}}}s_i(t)}{\lambda N}~,
 \label{Eq:rt}
\end{equation}
where $N$ is the number of traders in $\mathcal{N}$ and $\lambda$
measures the market depth or liquidity and is taken constant. In
expression (\ref{Eq:rt}), the decisions $s_i(t)$ are formed
slightly before $t$, in the period from $t-1$ to $t$ and are then
converted into price at $t$ by the market clearing process.

We account for the adaptive nature of agents and their learning
abilities by updating the coefficient of influence of agent $j$ on
agent $i$ according to the following rule:
\begin{equation}
K_{ij}(t) = b_{ij} +  \alpha_{i} K_{ij}(t-1) +  \beta r(t-1)
G(t-1) ~.
 \label{Eq:Kimemo}
\end{equation}
In this, we follow the large literature on the rationality of
imitation and of adaptive behavior when lacking sufficient
information or when this information seems unreliable \citep[see,
e.g.,][]{BrockHomme1,BrockHomme1bis,BrockHomme2,BrockHomme3,BrockHomme4,
BrockHomme5,Lebaronrev,Kirman,Lux1,Lux2,LuxMarchesi,LuxMarchesi2,Taka1,
Hubermancrash,Solomon1,Taka2,Solomon2,Gaunersdorfer}. The
coefficients $b_{ij}$ quantify the intrinsic imitation influence
of agent $j$ on agent $i$ in absence of other effects. For
$\alpha_i=\beta=0$, we recover a constant coefficient of
influence, which derives from a simple argument of bounded
rationality recalled in the Appendix with (\ref{jgfhha}). The
coefficient $\alpha_i>0$ (possibly different from one agent to
another) quantifies the progressive loss of memory of past
influences on the present. The last term with $\beta \neq 0$
quantifies how agent $i$ updates her propensity for imitation
based on the role of the exogenous news $G(t)$ in determining the
sign and amplitude of the observed return in the preceding time
period. This update depends upon whether the news $G(t-1)$ known
between time $t-2$ and $t-1$ has the same sign as the price
variations from $t-2$ to $t-1$, i.e., the same sign as the return
$r(t-1)$ defined by (\ref{ngmldl}). In addition to the sign, the
amplitude of the return is also taken into account in the updating
rule of the coefficient $K_{ij}(t)$ quantifying the propensity to
imitate: indeed, a small amplitude of the return has low
psychological as well as financial consequences and should not
count as much as a large amplitude of the return. The simplest
specification is to take into account the impact of the amplitude
of the return linearly in its size, hence the form $\beta r(t)
G(t-1)$ in (\ref{Eq:Kimemo}). Stronger nonlinear dependence is
probably more relevant \citep{PandeyStauffer,IS,SI} but is not
considered further here to keep the discussion simple.
\citet{prlrandis} have considered a simpler version in which each
$K_{ij}(t)$ is purely random and is constructed as the sum of two
random noises, one which is common to all coupling coefficients
and one which is specific to it. They are able to reproduce
volatility clustering and a power law distribution of returns at a
single fixed time scale.

\cite{G03AFM} considered this model (\ref{Eq:Kimemo}) with
$\alpha_i=0$, i.e., with no memory of past influence on present
influence, which leads the time series of $K_{ij}(t)$ looking like
a white-noise process. Our addition of the memory effect modifies
this white-noise structure into a Ornstein-Uhlenbeck type noise,
tending to a random-walk-like process for $\alpha_i=1$. With the
new term $\alpha_i \neq 0$, $K_{ij}(t)$ keeps a memory of past
successes that the news had on predicting the stock market moves
over approximately $1/(1-\alpha_{i})$ time steps. $\alpha_{i}$
thus characterizes the strength of the persistence of the links of
agent $i$ with other agents.

The sign of the coefficient $\beta$ is crucial.
\begin{enumerate}
\item For $\beta <0$, agent $i$ is less and less influenced by
other agents, the better has been the success of the news in
determining the direction and amplitude of the market return. This
process is self-reinforcing since, as $K_{ij}$ decreases, the
dominant term becomes $ \sigma_i(t) G(t)$, which further ensures
that the news correctly predict the decision of agents and
therefore the direction of the market move, thus decreasing
further the coefficient of influence $K_{ij}$. Reciprocally,
agents tends to be more influenced by others when the news seems
to incorrectly predict the direction of the market. The news being
not reliable, the agents turn to other agents, believing that
others may have useful information (see below and the Appendix for
an elaboration of this argument).

Taking a negative $\beta$ corresponds to agents behaving according
to standard rational expectations with respect to the flow of
external news. If the stock market is in agreement with the news
most of the time, a rational agent would conclude that the impact
of imitation, of herding, of trend following and of other
endogenously generated positive feedbacks, is minor and the news
are the dominating factor. Indeed, standard economics views the
stock market as a machine transforming news into prices and the
market is efficient when all news have been correctly incorporated
and are continuously incorporated into the market prices. In our
framework, this situation arises when the strategy
$s_i(t)={\rm{sign}}[G(t)]$ consisting in following the news is
found at least as good as or better than (\ref{Eq:Sit22}). The
agent would in this case decrease its propensity $K$ to imitate,
and continue to do so as long as the news are most of the time in
agreement with the stock market moves. We can thus summarize this
case  $\beta <0$ as describing ``boundedly rational'' agents.

\item For $\beta >0$, the more the news predict the direction of
the market, the more the agents imitate other agents. In other
words, there is a reinforcement of the influence between agents
when the news and the stock market return match at the previous
period. This is the ``irrational'' case where agents either
mis-attribute the origin of the market moves to herding rather
than to the impact of news, or misinterpret the exogenous
character of news in terms of endogenous herding or infer that
other agents will be following more eagerly as a group the
direction given by the news.

The regime $\beta>0$ may result from several mechanism.
\begin{itemize}
\item {\bf Mutually-reinforcing optimism}. When the market is
rising ($r(t-1)>0$) and the news are good ($G(t-1)=1$), the agents
may seek each other in order to determine if the rise can be
sustained: if the agent's neighbors are all bearish, then the
agent interprets this as a sign that the rise cannot be sustained
due to a lack of majority support; if the friends are bullish, the
agent is encouraged to feel optimistic, in the sense that more
good news may be on the way. It is thus not so much the price rise
or fall that agents try to predict and reflect but rather the
continuation of good news (or bad news) in the future and whether
they expect or not the good news to continue. Since a rise in the
market is the result of more people feeling optimistic that more
good news are on the way, then the optimism is spreading like an
epidemic and the market rises, reinforcing the influence
coefficient $K_{ij}$ (called propensity to be influenced by the
felling of others by \cite{G03AFM}).

\item {\bf Overconfidence}. Another mechanism is the tendency for
humans to exhibit overconfidence in their abilities, either
individually or as a group. If they see that the stock market has
moved in the same direction as the news indicated, they may
conclude that the information provided by the other agents has
been valuable, since the stock market is supposed to follow the
rule of the majority. Agents may thus be tempted to increase their
imitation behavior as long as news and stock market continue to
match. Or said differently, they attribute a larger value in the
prediction of the news which they interpret as being influenced by
the majority opinion. Indeed, experiments committed by
\cite{Darkefreedman} show that a lucky event can lead to
overconfidence. \cite{heathgonz} have also compared decisions made
alone to decisions made following interactions with others, and
shown that, while interaction did not increase decision accuracy
or meta-knowledge, subjects frequently showed stable or increasing
confidence when they interacted with others, even with those who
disagreed with them \citep[see
also][]{Savoybeitel,Slaterrouner,robertscarl,Ropbythor}. A
possible interpretation is that the interaction serves the role of
rationalizing the subjects' decisions rather than collecting
valuable information. In the same spirit, \citet{Sieck} have shown
that exposition to others of the rationale behind decisions
increase markedly subjects' confidence that their choices were
appropriate.

\end{itemize}

\end{enumerate}

In summary, the model contains the following general ingredients:
(i) the agents make decisions based on a combination of three
ingredients: imitation, news and private information; they are
boundedly rational; (ii) traders are heterogeneous ($K_{ij}$ and
$\sigma_i$); (iii) The propensity to imitate and herd is evolving
adaptively as an interpretation that the agents make of past
successes of the news to predict the direction of the market.

\subsection{Specification of the updates of expectations of
other agents' decisions \label{mgmlmsls}}

The model is completely specified once the algorithm, used to
construct how an agent estimates her expectation ${\rm E}[s_j](t)$
of other agents' decisions in  expression (\ref{Eq:Sit22}), is
given. Three possibilities can be considered.
\begin{itemize}
\item ${\rm E}[s_j](t) = s_j(t-1)$: agent $i$ expects a
persistence of the other agents' decisions, similarly to a
martingale condition; in absence of any information other than the
past actions, the next predicted action is the last one. This
leads to transform (\ref{Eq:Sit22}) into the prescription
\begin{equation}
 s_i(t)={\rm{sign}}\left[\sum_{j\in{\mathcal{N}}}K_{ij}(t)s_j(t-1)+
 \sigma_i(t) G(t) +\epsilon_i(t)\right]~.
 \label{Eq:Sit}
\end{equation}
Without the term $\sigma_i(t) G(t)$, this was the model adopted by
\cite{JLS00IJTAF} to develop a theory of herding to explain
financial bubbles as regimes of strong imitation between agents.
\citet{Kaizoji} have also studied the
dynamics of a stock market with heterogeneous agents in
the framework of a recently proposed spin model for the emergence of
bubbles and crashes.
The Appendix recalls the derivation by \cite{RS00} showing how
this specification is a natural consequence of bounded rationality
of agents.

\item ${\rm E}[s_j](t) = s_j(t)$: each agent $i$ has access to the
information of the action of other traders instantaneously and she
can not do better than use it. Computationally, the solution ${\rm
E}[s_j](t) = s_j(t)$ is self-consistent since $s_i(t)$ depends on
$s_j(t)$ which itself depends on the former. Such self-consistent
formulation can be solved by using an iterative algorithm as
follows
\begin{equation}
{\rm E}[s_j](t) = \lim_{k\to\infty}s_j^{[k]}(t)~, \label{Eq:Esi}
\end{equation}
 where
\begin{equation}
 s_i^{[k+1]}(t)={\rm{sign}}\left[\sum_{j\in{\mathcal{N}}}K_{ij}(t) s_i^{[k]}(t)+
 \sigma_i(t) G(t) +\epsilon_i(t)\right]
 \label{Eq:Si2Iter}
\end{equation}
with, for instance, the starting condition
$s_i^{[0]}(t)=s_i(t-1)$.

\item Information cascade along a given path $i=1,2,\cdots,N$
where the ordered sequence of indices $i$ encodes the curvilinear
abscissa along a path linking all agents, in the spirit of the
information cascades \citep{Welch1992,Bikhchandani}. In such a
scheme, ${\rm E}[s_j](t) = s_j(t-1)$ if $j > i$ and ${\rm
E}[s_j](t) = s_j(t)$ if $j < i$ when considering agent $i$. In
other words, we have
\begin{equation}
 {\rm E}[s_j](t)=H(j-i) s_j(t-1)+H(i-j) s_j(t)~,
 \label{Eq:Si3Iter}
\end{equation}
where $H$ is the Heaviside function.
In the literature on adaptive learning, this procedure is
said to be backward-looking, because it does not model the other agents'
future behavior given all past information, for instance, because agents
are not able to form beliefs about the other agents' future choices.
These rather naive precedures have been suggested by psychologists and
animal behaviorists in the 1950s and have been tested in laboratory
settings.

\end{itemize}

\subsection{Restriction to model \citep{G03AFM} with $\alpha_i=0$
corresponding to the absence of memory of the coefficients
$K_{ij}$'s of imitation \label{mgmlls}}

Our model is a straightforward generalization of the artificial
stock market model formulated by \cite{G03AFM}, in which the
coefficients $K_{ij}$ of the influence of $j$ on $i$ are taken
identical for all $j$'s and are updated as follows:
\begin{equation}
 K_{ij}(t) = K_i(t) = b_i + \beta r(t-1) G(t-1)~.
 \label{Eq:ASMKi}
\end{equation}
The coefficients $b_i$'s capture the ``natural'' propensity of
humans for imitation, which may vary from agent to agent. There
are drawn at the beginning of the simulation from a uniform
distribution between $0$ and some maximum positive value and
remain fixed thereafter during the dynamics. The main difference
between (\ref{Eq:ASMKi}) and our specification (\ref{Eq:Kimemo})
is the absence of the persistence or memory of $K_{ij}(t)$ on its
past values in (\ref{Eq:ASMKi}).

\begin{figure}
\begin{center}
\includegraphics[width=14cm]{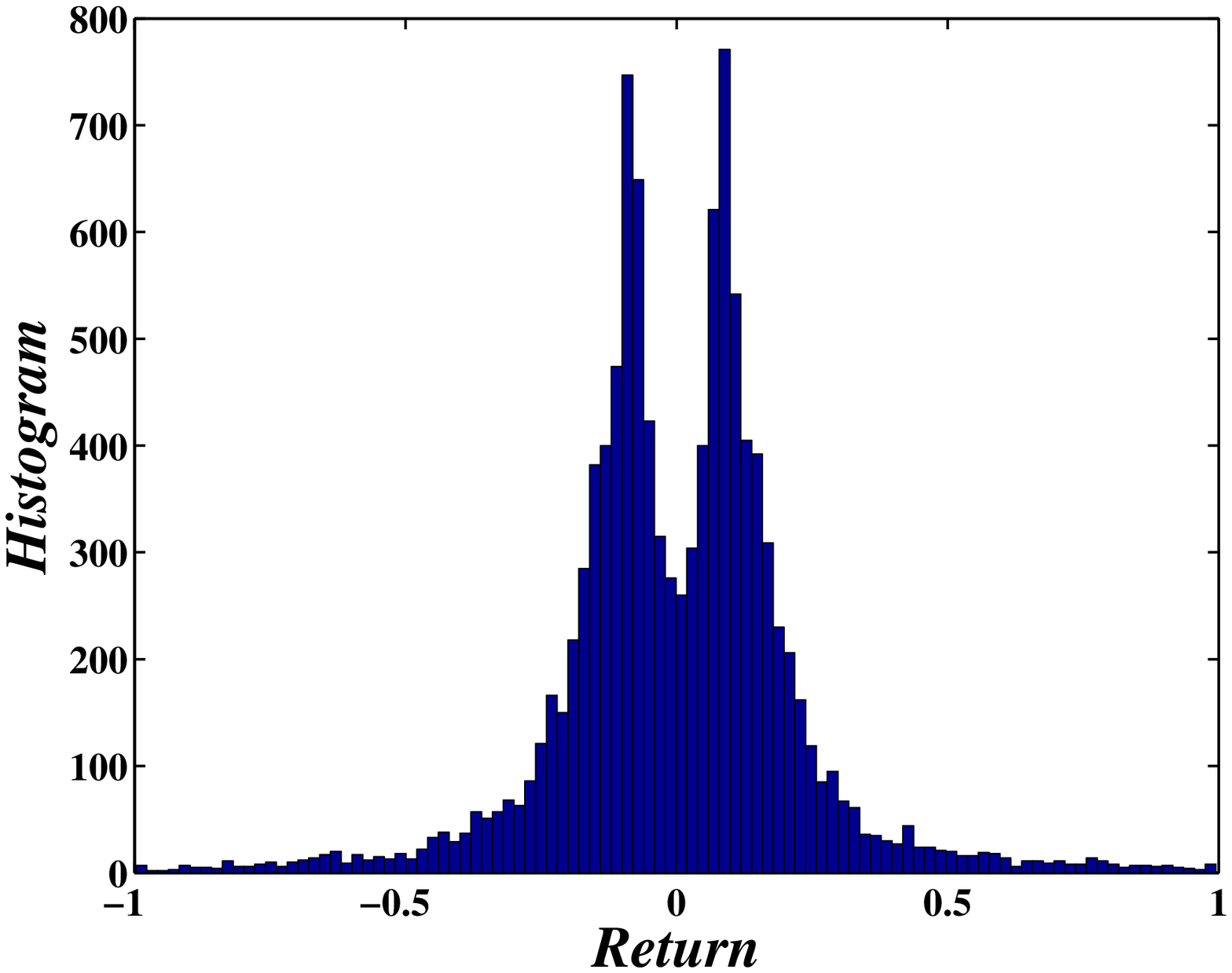}
\end{center}
\caption{Density distribution of returns $r_1$ for a realization
of the artificial stock market model formulated by \cite{G03AFM}
generated using $b_{\max} = 0.22\sim 0.24$, $\sigma_{\max} = 0.14
\sim 0.15$ and $CV = 0.8 \sim 0.9$ as recommended by this author.
The time series of returns have been kindly provided by
Gon\c{c}alves. Our own simulations reproduce the same results.}
\label{Fig:ReturnHist:Carlos}
\end{figure}

Based on numerical simulations and synthesis, \cite{G03AFM} argues
that the model (\ref{Eq:Sit22}) using the implementation
(\ref{Eq:Si3Iter}) corresponding to an information cascade along a
specific path and with (\ref{Eq:ASMKi}) reproduce all the
important stylized facts of the stock market only for $\beta > 0$.
This is interesting but actually not quite correct. Consider, for
instance, the following parameters $b_{\max} = 0.22\sim 0.24$,
$\sigma_{\max} = 0.14 \sim 0.15$ and $CV = 0.8 \sim 0.9$, which
are recommended by \cite{G03AFM} to reproduce the main stylized
facts (fat tailed distributions of returns, clustering of
volatility, bubbles, crashes). The problem is that the time series
of returns generated with these parameters have unrealistic
bimodal distributions of returns, as shown in
Fig.~\ref{Fig:ReturnHist:Carlos}. The origin of this bimodal
structure results from the fact that the model explores the
ordered regime of the Ising model too often, corresponding roughly
speaking to an average coupling constant $\langle K_i \rangle$
larger that the critical Ising value $K_c$. In this case, agents
form a crowd with a majority opinion which can flip with time
between $\pm 1$ due to the feedback process of the news and return
dynamics on the coupling coefficient. In other words, the two
bumps of the distribution of returns are direct signatures of the
existence of the two spontaneous ordered states of the Ising model
when the coupling coefficient is above its critical value. The
dynamics of decisions is thus characterized by more or less random
flips of crowds of agents between the two opinions. The resulting
stylized facts can not therefore be taken as genuine signatures of
a realistic dynamics of agents' opinions.

\section{Simulations and results of the model with memory in the
dynamics of influence coefficients}

The origin of the unrealistic bimodal distribution of returns
shown in Fig.~\ref{Fig:ReturnHist:Carlos} has to be found in the
rather artificial property that the influence coefficients
$K_{ij}(t)$'s, which are updated according to (\ref{Eq:ASMKi}),
lose instantaneously from one time to the next the memory of their
past values. As a consequence, the coefficients $K_{ij}(t)$
fluctuate with time approximately as a white noise with an
amplitude controlled by that of the returns and this gives rise to
abrupt shifts of the majority opinion as explained in the previous
section. This property is unrealistic to model the real
persistence of interactions between agents. Indeed, in the real
world, people are connected through social networks that do not
instantaneously reshuffle in response to external effects as
expression (\ref{Eq:ASMKi}) describe. Rather, social connections
evolve slowly and exhibit significant persistence, as documented
in numerous studies \citep[see, for instance,][and references
therein]{Suitor,Wellman}. Networks of investors communicating
their opinions and sentiment on the stock market are similarly
persistent.

We thus turn to model (\ref{Eq:Sit22}) implemented with
(\ref{Eq:Sit}) or along a specific information cascade
(\ref{Eq:Si3Iter}), as described in section \ref{mgmlmsls} We
assume the existence of a memory in the dynamics of the influence
coefficients, which are updated according to the simplified
version of (\ref{Eq:Kimemo}) given by
\begin{equation}
K_{i}(t) = b_i +  \alpha K_{i}(t-1) + \beta r(t-1) G(t-1) ~.
 \label{Eq:RevASM:Ki}
\end{equation}
where the $b_i$'s are uniformly distributed in $(0,b_{\max})$ at
the beginning of the simulation and remain frozen thereafter. The
value $b_i$ of agent $i$ represents her idiosyncratic imitation
tendency. Compared with (\ref{Eq:Kimemo}), the memory parameter
$\alpha$ is taken the same for all agents. We use a $50\times 50$
lattice as the geometrical implementation of the social network,
in which the agents are located at the nodes and each agent
interacts only with her four nearest neighbors. The sensitivity
$\sigma_i$ of agent $i$ to the global news $G(t)$ is uniformly
distributed in the interval $(0,\sigma_{\max})$. The value
$\sigma_i$ of agent $i$ is again specific to her and quantifies
her susceptibility to be influenced by external news. The
idiosyncratic or private information term $\epsilon_i(t)$ of agent
$i$ is drawn at each time step from a normal distribution with
zero mean and a standard deviation $s_{\epsilon,i}$ which is also
different from one agent to another: $s_{\epsilon,i}$ is chosen at
the beginning of the simulation (like $b_i$ and $\sigma_i$) to
characterize agent $i$ according to a value equal to the sum of a
common constant $CV$ and of a uniform random variable in the
interval $[0,0.1]$.

In our simulations, we fix $\lambda =40$ (which determines the
scale of the returns to a value comparable to that of empirical
observations) and $\alpha=0.2$. We have also investigated other
values $\alpha=0.4$, $0.6$, and $0.8$ and obtain similar results.
We explore the properties of the model in the parameter space of
$b_{\max}$, $\sigma_{\max}$ and $CV$. There is no loss of
generality in fixing $|\beta|=1$ to explore the relative
importance of the term $\beta r(t-1) G(t-1)$, since the typical
scale of the $K_i$'s is set by $b_{\max}$ whose amplitude is
varied in our numerical exploration. However, the sign of $\beta$
is fundamental as explained in section \ref{jfjs}. We thus
consider in turn the two cases $\beta=-1$ and $\beta=+1$ and
explore in each case a large sample of triplets $(b_{\max},
\sigma_{\max}, CV)$.

We ask whether the model can account for the most often reported
stylized facts of financial markets. In other words, we would like
to validate the model. For this, we consider the following
metrics: (i) the distribution of returns at different time scales;
(ii) the correlation function of returns and of the absolute value
of the returns (taken as a proxy for the financial volatility);
(iii) the scaling of the moments of increasing orders of the
absolute values of the returns (testing multifractality); (iv) the
existence of a hierarchy of exponents controlling the relaxation
of the volatility after an endogenous shock (another hallmark of
multifractality); (v) the existence of bubbles and crashes and
their properties. Our strategy is to search for a robust set of
model parameters for which all these stylized facts are reproduced
not only qualitatively but also quantitatively. The main result of
our analysis is that it is impossible to validate the model for
$\beta<0$ while we find sets of parameters for $\beta>0$ which
nicely fit the stylized facts of real financial markets.

\subsection{News predicts the next return $\to$ decrease of imitation: $\beta=-1$}
\label{s4:REmodel}

In this case, when the news $G$ happen to correctly predict the
return ($rG>0$), the agents reduce their mutual imitation. On the
contrary, when $rG<0$, imitation among agents strengthens. Thus,
$\beta<0$ corresponds to agents behaving according to standard
rational expectations with respect to the flow of external news: a
rational agent would concluded that, if the stock market is in
agreement with the news then, the impact of imitation is minor and
the news are the dominating factor, as standard economic textbook
describe, since the strategy $s_i(t)={\rm{sign}}[G(t)]$ consisting
in following the news is found as good as or better than the more
elaborate dynamics of the agents' actions $s_i(t)$ incorporating
the imitation process. For $\beta <0$, agents decrease their
propensity to imitation as long as the news are in agreement with
the stock market moves.

The following argument shows that the attractor of the dynamics is
characterized by negligible imitation and only the news and
private information terms are important for the dynamics. Consider
a population of traders at time $t$ with their propensity $K_i(t)$
to imitate on average above the critical Ising value $K_c$ such
that imitation initially dominates the dynamics. The news $G(t)$
become known and the decisions of the agents given by
(\ref{Eq:Sit22}) decide collectively the return $r(t+1)$ from time
$t$ to $t+1$ through expression (\ref{Eq:rt}). Since the
$K_i(t)$'s are overall above $K_c$, it is well-known from many
past studies of the Ising model (see for instance \cite{Ising1}
and references therein) that the corresponding ``ferromagnetic''
phase of the system is characterized by a strong slaving to
external fields such as $G(t)$. Hence, the collective opinion
$\sum_i s_i$ takes the sign of $G(t)$ with a high probability. As
a result, with a large probability, $r(t+1) G(t)$ is positive,
which entails a downgrading of $K_i$ by an amount $r(t+1)$ since
$beta=-1$ (in addition to the other terms which tend to reverse
$K_i(t)$ to the value $b_i/(1-\alpha)$). This behavior continues
as long as the $K_i$'s are above or close to $K_c$ (since the
$K_i$'s are heterogeneous, the effective critical values is
modified compared with the homogeneous case and our argument
remains valid when using this modified value). Alternatively, if
the $K_i$'s are on average smaller than $K_c$, the collective
decision $\sum_i s_i$ and therefore the market return have little
or no relationship with the external news. Hence, the term $\beta
r(t+1) G(t)$ takes random signs from one time step to the next,
leading to a effective random forcing added to the autoregressive
equation $K_i(t) = b_i + \alpha K_i(t-1)$. The coefficients
$K_i(t)$ evolve to fluctuate around the asymptotic value
$b_i/(1-\alpha)$. We thus expect Gaussian distributions of returns
when $b_i/(1-\alpha)$ is smaller than $K_c$ and bimodal
distributions when $b_i/(1-\alpha) > K_c$ reflecting the slaving
of the global opinion to the sign of the news.

In our simulations, we have scanned $b_{\max}$ from $0.1$ to $0.5$
with spacing $0.1$, $\sigma_{\max}$ from $0.005$ to $0.08$ with
spacing $0.005$, and $CV$ from $0.1$ to $1.1$ with spacing $0.2$.
This corresponds to a total of 480 different models. We use
(\ref{Eq:Sit22}) implemented according to the information cascade
of sentiment formation explained in the item associated with
equation (\ref{Eq:Si3Iter}). This choice is made to minimize the
computational cost. Tests using the other updates over limited
time span suggest that the results we are interested in are not
sensitive to the details of the updating rules. We run each model
over $10^4$ time steps and then analyze the time series of
returns.

\begin{figure}
\begin{center}
\includegraphics[width=14cm]{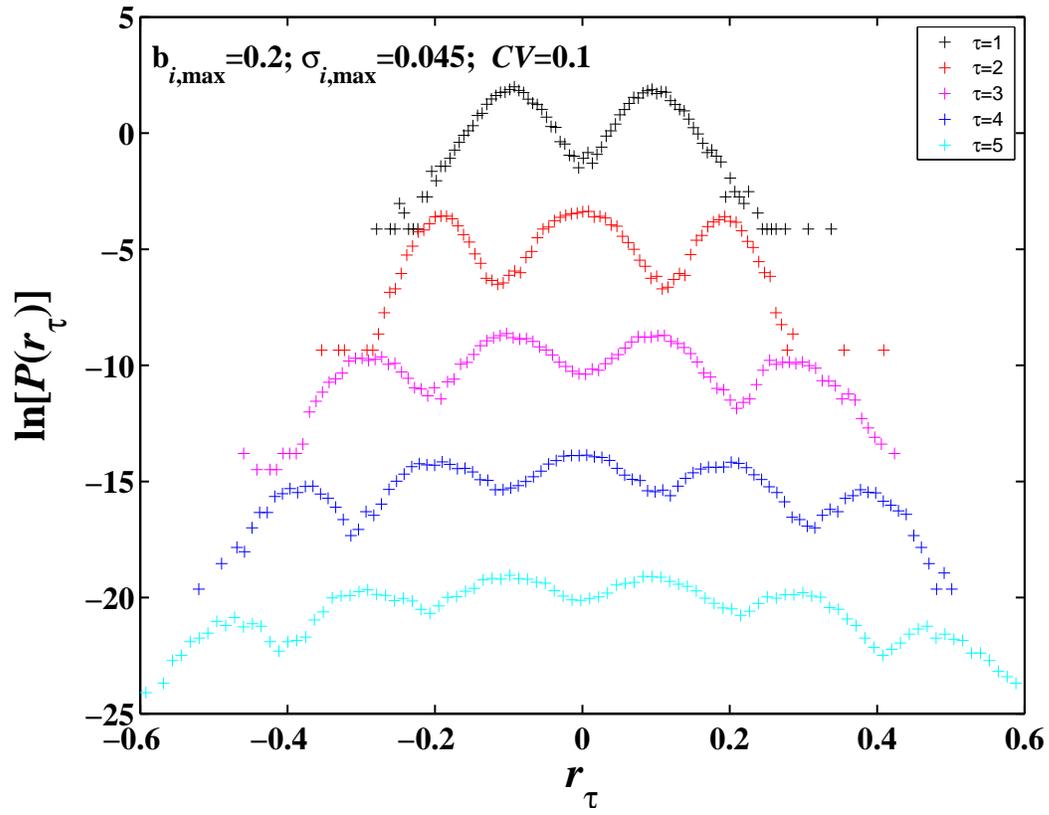}
\end{center}
\caption{A typical example of the multimodal distribution for
$b_{\max}=0.2$, $\sigma_{\max}=0.045$, and $CV=0.1$.}
\label{Fig:Bimodal}
\end{figure}

The first metric we analyze is the distribution of returns at
different time scales $\tau$, defined according to
\begin{equation}
 r_{\tau}(t) = \ln[p(t)/p(t-\tau)]~.
 \label{Eq:rtau}
\end{equation}
where $\tau$ is a multiple of the time step. We observe two
classes of shapes. For large idiosyncratic noise (large $CV$) and
not too large $b_{\max}$, the distribution of returns is Gaussian
for all time scales $\tau$. For smaller $CV$'s and larger
$b_{\max}$, we observe multimodal return distributions, as
illustrated by the typical example shown in Fig.~\ref{Fig:Bimodal}
obtained for $b_{\max}=0.2$, $\sigma_{\max}=0.045$, and $CV=0.1$.
The number of peaks in the distribution of $r_{\tau}$ is $\tau+1$.
These multimodal distributions correspond to the regime where the
news $G(t)$ controls the collective opinion of the traders which
tend to coordinate their decisions as explained above when
$b_i/(1-\alpha)$ is larger than $K_c$. For given $b_{\max}$ and
$CV$, the bimodal structure of the distribution of $r_1$ becomes
more and more significant as $\sigma_{\max}$ increases.
Alternatively, for fixed $\sigma_{\max}$ and $CV$ (say,
$\sigma_{\max}=0.02$ and $CV=0.1$), a bimodal distribution is
obtained for sufficiently small $b_{\max}$ and we observe a
crossover from bimodal ($b_{\max}=0.1$) to unimodal
($b_{\max}=0.5$) with a crossover with a plateau ($b_{\max}=0.3$).
This corresponds to the regime in which a majority of agents react
to the global news in the same manner and buy or sell
simultaneously. Since the news are $G(t) = \pm 1$ with equal
probability, the decisions of the agents and therefore the returns
most often jump between two values of equal amplitude and opposite
signs. The multi-modal structure of the distributions of returns
at time scale $\tau$ then results from the properties of the
convolution of the bimodal distribution of the returns $r_1$.

In the parameter space that we have explored and notwithstanding
our best attempts, we have not been able to find a set of
parameters leading to distributions of returns exhibiting a
monomodal shape with fat tails for small time scales, evolving
slowly towards Gaussian distributions at large time scales, as can
be observed in empirical data
\citep{Ghashghaie-1996-Nature,Mantegna-Stanley-2000}.
% This is given in GASM_Sim_05.mat.

In addition, we observe that the correlation function of returns
($C_{\tau}(r,r)$) and of volatilities ($C_{\tau}(|r|,|r|)$) have
similar amplitudes and decay with the same characteristic time
scale as a function of time lag. This is very different from the
observed correlations of financial markets, with very short memory
for returns and long-memory for the volatility.

We have also investigated the impact of the updating rule of the
agents' sentiments. If instead of the information cascade (third
item of section \ref{mgmlmsls}), we use the parallel update
(\ref{Eq:Sit}), for the same range of parameters $b_{\max}$,
$\sigma_{\max}$, and $CV$, we find that most of the returns are
between two values proportional to $\pm 1$ and the distribution of
returns is close to $P(r_1)=\delta(r_1^2-1)$. In other words, the
bimodality of the distribution of returns is much more pronounced
than shown in Fig.~\ref{Fig:Bimodal} that was obtained for the
information cascade updating scheme. For large idiosyncratic noise
(large $CV$), the bimodality disappears and is replaced by pdf's
which are approximately Gaussian. With $CV$ and $b_{\max}$ (resp.
$\sigma_{\max}$) fixed, the volatility increases with increasing
$\sigma_{\max}$ (resp. decreasing $b_{\max}$) and the distribution
is bimodal when $\sigma_{\max}$ (resp. $b_{\max}$) is large (resp.
small) enough. The time evolution exhibits in addition long
transients with returns fluctuating around zero before bifurcating
to the bimodal state. The duration of the transient decreases with
increasing $\sigma_{\max}$ or $b_{\max}$. Scanning the parameters,
we never obtained realizations with realistic distributions of
returns at different time scales. We conclude that the updating
scheme (\ref{Eq:RevASM:Ki}) with $\beta=-1$ corresponding to
boundedly rational agents cannot explain the stylized facts of
empirical finance.

\subsection{News predicts the next return $\to$ increase of imitation: $\beta=1$}
\label{s3:OCmodel}

In our simulations, we fix $\alpha=0.2$, $b_{\max}$ varies from
$0.1$ to $0.5$ with spacing $0.1$, $\sigma_{\max}$ from $0.01$ to
$0.08$ with spacing $0.01$, and $CV$ from $0.1$ to $0.7$ with
spacing $0.2$. This gives 160 models that we explore by generating
time series of length equal to 10000 time steps. We use
(\ref{Eq:Sit22}) implemented according to the information cascade
of sentiment formation explained in the item associated with
equation (\ref{Eq:Si3Iter}).
% This is given in GASM_Sim_01.mat.
% WXZHOU_StylizedFacts(cumsum(r{3,3,1}), 'GASM_OC', num);
%   We redid the simulation with longer time evolution to have better
% statistics for the parameters (0.3,0.03,0.1) used below and the
% simulated returns are stored in GASM_Sim_OC_bubble.mat.

When $CV$ is very large, the distribution of return $r_{\tau}$ is
Gaussian, simply because the returns are dominated by the
idiosyncratic noise modeling private information, which is chosen
Gaussian. This regime is not interesting since it erases both the
effect of news and of imitation. For smaller $CV$'s, we also
observe bimodal distributions of the returns $r_1$ for certain
ranges of parameters, as in the case $\beta=-1$. This occurs for
large $\sigma_{\max}$ and small $b_{\max}$. Similar bimodal
distributions of returns are obtained for $\alpha=0$, as described
in section \ref{mgmlls}.

We have found several parameter combinations which lead to
realistic stylized facts. For instance, for the following sets of
$(b_{\max}, \sigma_{\max}, CV)$, (0.3, 0.03, 0.1), (0.4, 0.04,
0.1), (0.4, 0.05, 0.1), (0.5, 0.06, 0.1), (0.1, 0.01, 0.3), (0.1,
0.02, 0.3), (0.2, 0.02, 0.3), (0.2, 0.03, 0.3), (0.3, 0.04, 0.3),
(0.5, 0.05, 0.3), (0.5, 0.07, 0.3), (0.3, 0.03, 0.5), and (0.5,
0.05, 0.5), the distribution of returns $r_{\tau}$ is a stretched
exponential (or close to a power law) \citep{Malpissor} for small
$\tau$, exponential for intermediate $\tau$, and Gaussian for
large $\tau$, as in real financial data.

In the following, we exemplify the obtained stylized facts with
$\alpha=0.2$ and with the parameter combination ($b_{\max}=0.3,
\sigma_{\max}=0.03, CV=0.1$) which is typical.

\subsubsection{Probability density functions of log-returns at different time scales}

\begin{figure}
\begin{center}
\includegraphics[width=14cm]{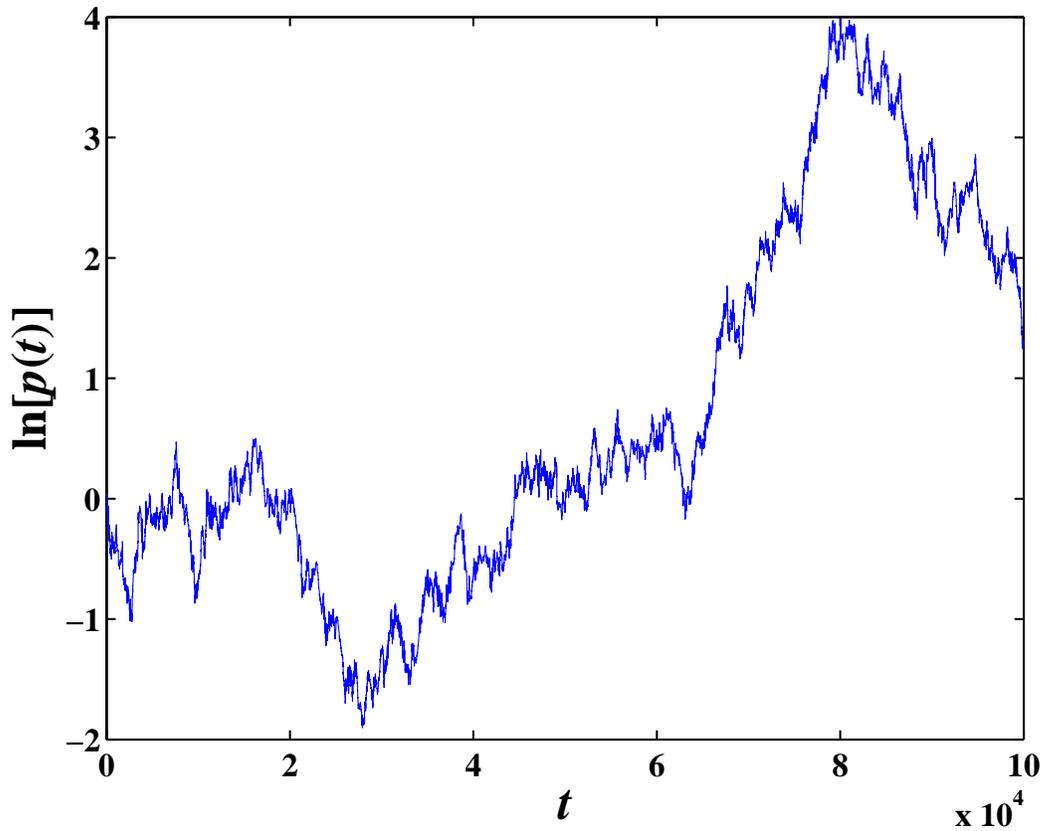}
\end{center}
\caption{A realization of the logarithm of the price over $10^5$
time steps generated using $\alpha = 0.2$, $b_{\max} = 0.3$,
$\sigma_{\max} = 0.03$ and $CV = 0.1$ of the generalized
artificial stock market model defined by (\ref{Eq:Sit22}),
(\ref{Eq:rt}) and (\ref{Eq:ASMKi}). } \label{Fig:OC:lnP}
\end{figure}

\begin{figure}
\begin{center}
\includegraphics[width=14cm]{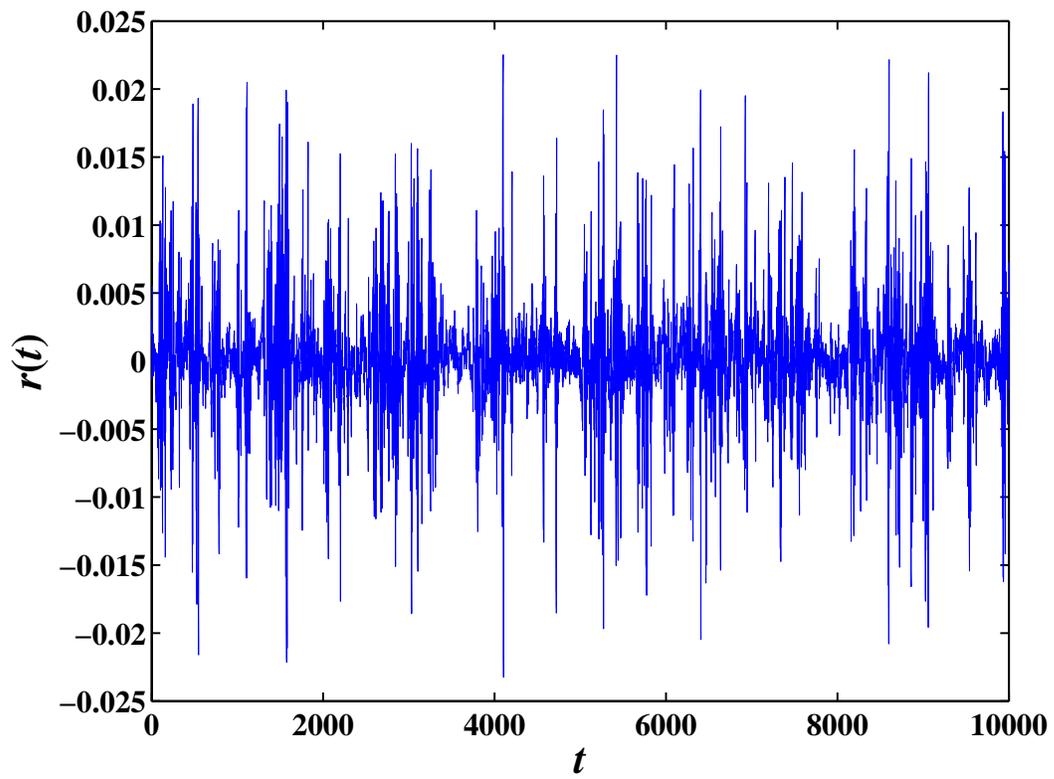}
\end{center}
\caption{Time series of the log-returns of the price shown in
Fig.~\ref{Fig:OC:lnP}. } \label{Fig:OC:R}
\end{figure}

\begin{figure}
\begin{center}
\includegraphics[width=14cm]{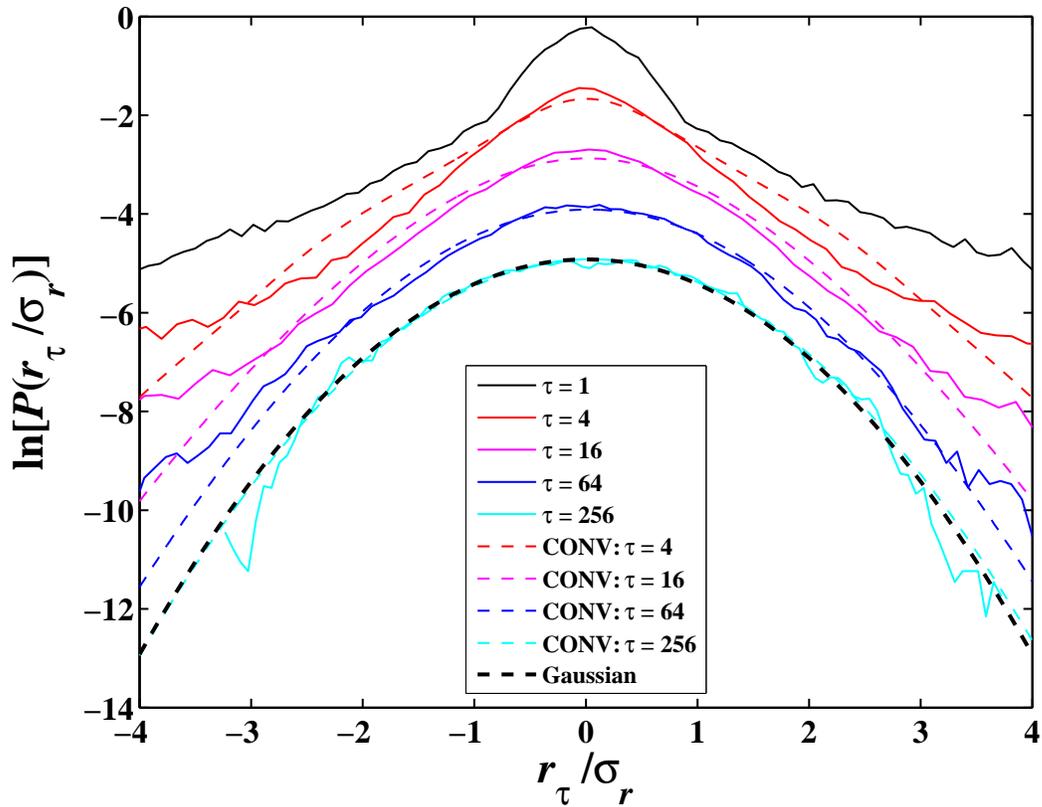}
\end{center}
\caption{(Color online) Empirical (solid lines) and theoretical
(dashed thin lines) probability distribution density (in
logarithmic scales) of log-returns at different time scales $\tau$
of the price time series shown in Fig.~\ref{Fig:OC:lnP}. The
log-returns $r_{\tau}$ are normalized by their corresponding
standard deviations $\sigma_{\tau}$. The pdf curves are translated
vertically for clarity. The thick dashed line is the Gaussian
pdf.} \label{Fig:OC:pdfR}
\end{figure}

Figure \ref{Fig:OC:lnP} shows a realization of the logarithm of
the price over a time interval of $10^5$ time steps. Figure
\ref{Fig:OC:R} shows the corresponding time series of the
log-returns defined by (\ref{ngmldl},\ref{Eq:rt}), where
$\lambda=40$ is set to scale the returns to realistic values. Note
the existence of clusters of volatility which are qualitatively
similar to those observed in real financial data. The solid lines
in Fig. \ref{Fig:OC:pdfR} show the logarithm of the probability
distribution densities (pdf) of the log-returns at different time
scales defined by (\ref{Eq:rtau}) as a function of the returns
$r_{\tau}$ scaled by their standard deviation $\sigma_{\tau}$. The
pdf curves have been translated vertically for clarity. In the
semi-log representation of Figure \ref{Fig:OC:pdfR}, a straight
line qualifies an exponential law. We observe stretched
exponential laws at short time scales that cross over smoothly to
a Gaussian law at the largest shown time scale. This evolution of
pdf's with time scales comply with the well-known stylized fact of
financial markets \citep{Ghashghaie-1996-Nature}. The model thus
obtains price series with the correct monomodal shape with fat
tails, and the correct progressive transition to a Gaussian
distribution at large time scales. We stress that this comparison
improves on those involving only one time scale.

It is interesting to compare the obtained pdf's at time scales
$\tau$ larger than the elementary time step $1$ with those which
would derive by $\tau$-fold convolution of the pdf at the unit
time scale. In absence of time dependence, the two should be
asymptotically identical. For instance, the pdf ${\rm{pdf}}(r_4)$
of returns $r_4$ over four time steps would be given by
${\rm{pdf}}(r_4)={\rm{pdf}}(r_1)\otimes {\rm{pdf}}(r_1)\otimes
{\rm{pdf}}(r_1)\otimes {\rm{pdf}}(r_1)$. Upon convolution in the
absence of dependence, the variance is additive which gives a
prediction for the standard deviation of the pdf of returns at
$\tau$ time steps, $\sigma_{\tau} = \sigma_1 \sqrt{\tau}$, which
can be used to normalize the pdfs in terms of the reduced variable
$r_{\tau}/\sigma_{\tau}$. The corresponding pdf's of $r_4$,
$r_{16}$, $r_{64}$, and $r_{256}$ obtained by convolution of the
pdf of $r_1$ are drawn in Fig.~\ref{Fig:OC:pdfR} with dashed
lines. We see clearly that the true pdf's for $\tau=4, 16, 64$
have a much fatter tail compared with the theoretical pdf's based
on absence of dependence. Such behavior is very similar to what is
observed in real data \citep[see for instance][Fig.
2.2]{Malsorbook}.

\subsubsection{Autocorrelations of log-returns and volatility \label{ngjs}}

\begin{figure}
\begin{center}
\includegraphics[width=14cm]{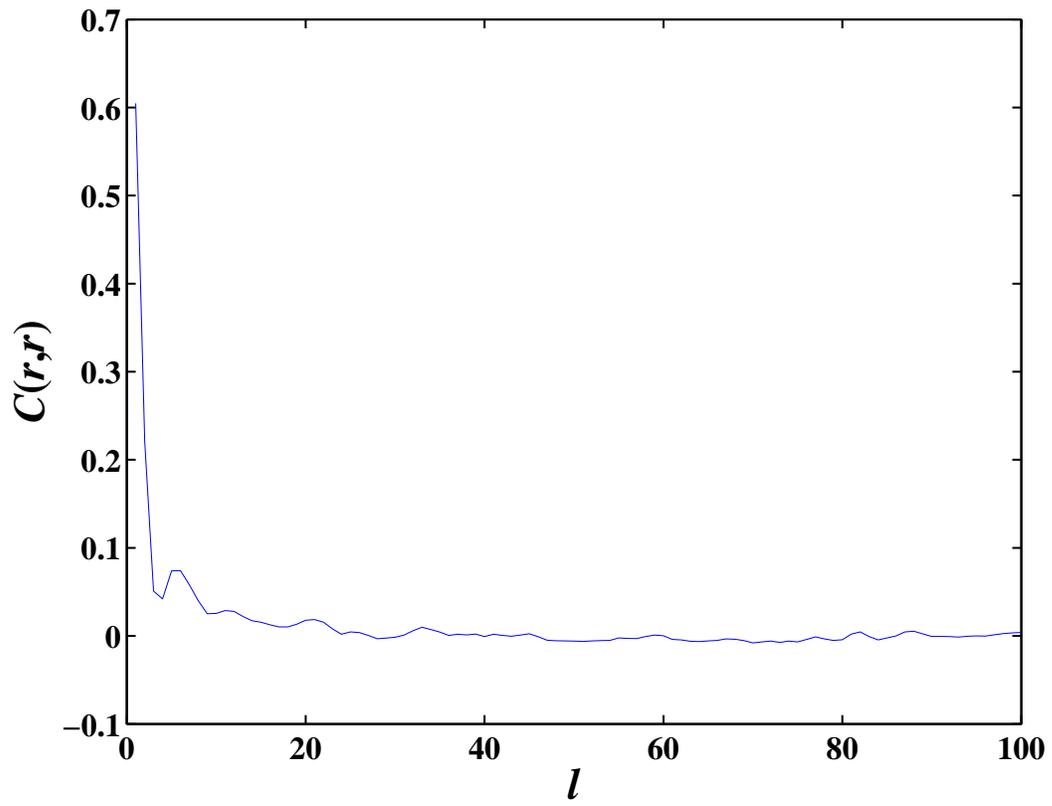}
\end{center}
\caption{Autocorrelation function of the log-returns of the
realization shown in Fig.~\ref{Fig:OC:lnP}. } \label{Fig:OC:CorrR}
\end{figure}

\begin{figure}
\begin{center}
\includegraphics[width=12cm]{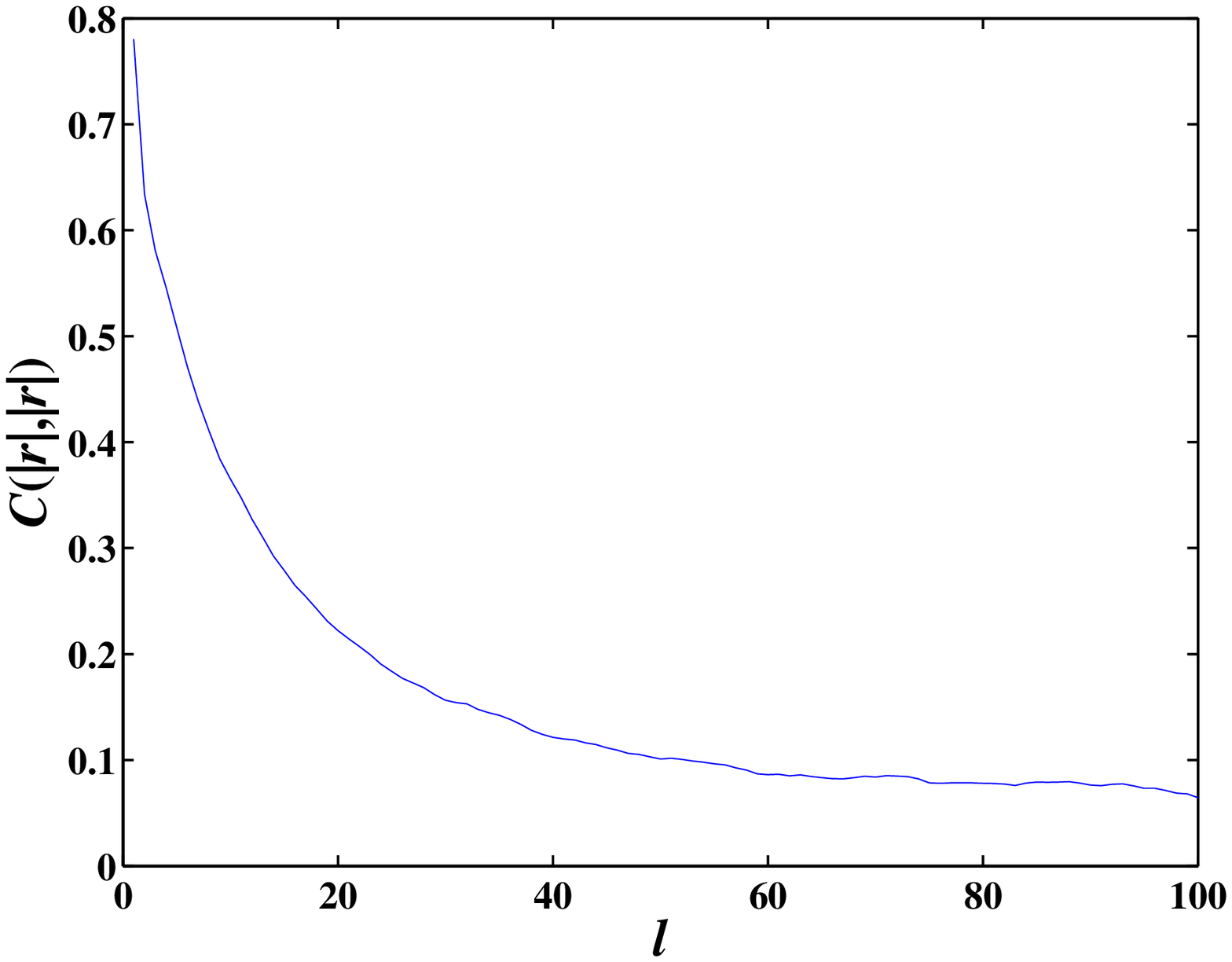}
\end{center}
\begin{center}
\includegraphics[width=12cm]{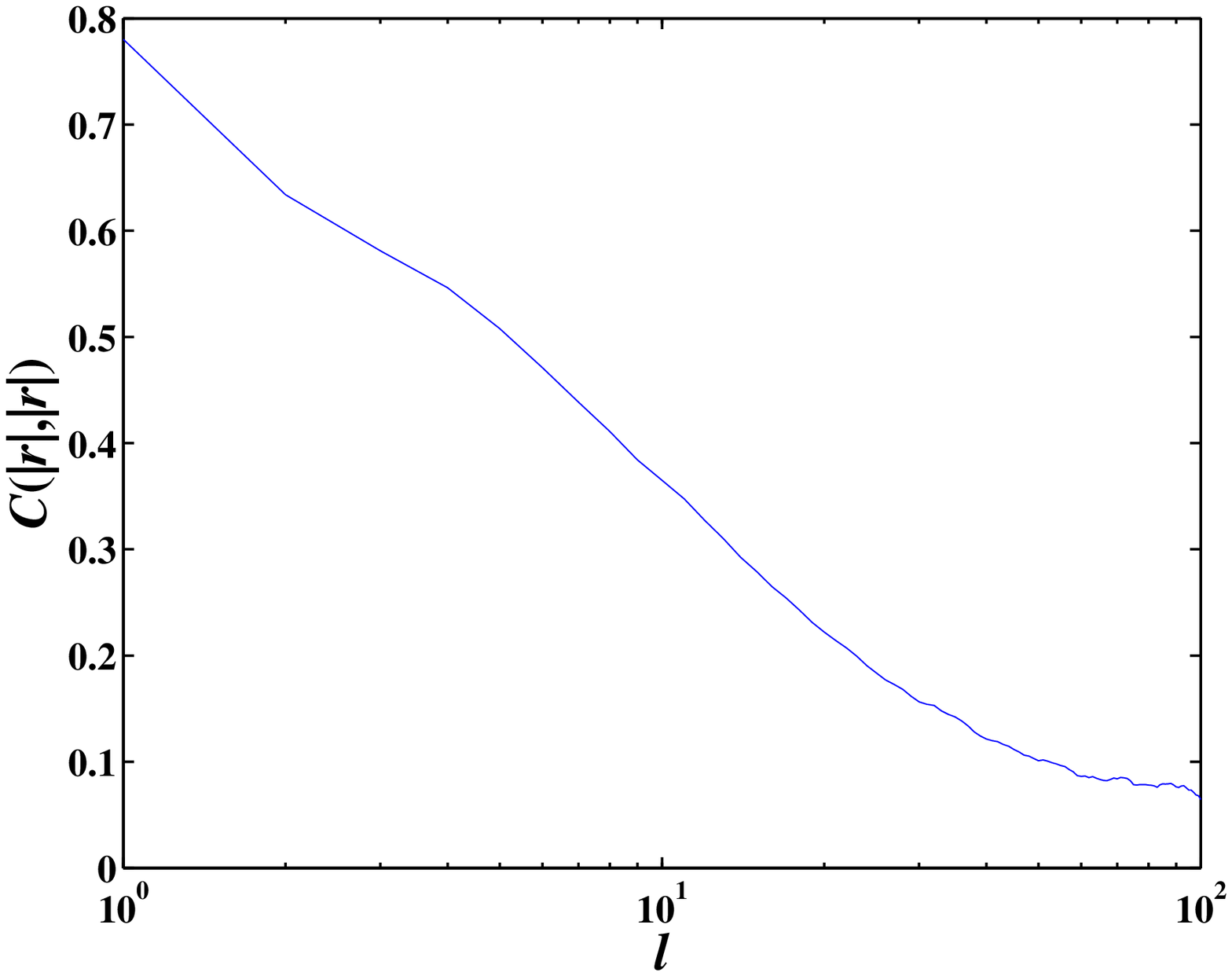}
\end{center}
\caption{Autocorrelation function of the absolute value of
log-returns of the realization shown in Fig.~\ref{Fig:OC:lnP}. The
top panel show the correlation in linear-linear scale. The bottom
panel plots the correlation function as a function of the
logarithm of the time lag, as suggested by the multifractal random
walk model (see text).} \label{Fig:OC:CorrAbsR}
\end{figure}

Figure \ref{Fig:OC:CorrR} shows the temporal correlation of the
log returns $r_1$ as a function of the time lag $\ell$. One can
observe a very short correlation time, of duration smaller than
one time step. Figure \ref{Fig:OC:CorrAbsR} presents the temporal
correlation of the absolute value of log returns $r_1$, taken as a
proxy for the volatility, both in linear-linear and in linear-log
scales. It is apparent that the volatility exhibits a strong
correlation with memory lasting approximately 100 time steps for
this set of parameter. The right panel of Figure
\ref{Fig:OC:CorrAbsR} represents the correlation of the volatility
as a function of the logarithm of the time lag. This
representation is suggested by the multifractal random walk (MRW)
model which is constructed by definition with a correlation
decaying linearly with the logarithm of the time lag, up to a
so-called integral time scale $T$
\citep{M_etal,B_etal,MuzyQF,endovol}. The right panel of Figure
\ref{Fig:OC:CorrAbsR} shows that this dependence suggested by the
MRW provides a reasonable approximation of the numerical data. The
integral time scale is here estimated around 100 time steps. These
observations are is good agreement with the stylized facts on the
correlation of returns and of volatility of real financial
markets. Indeed, one of the key stylized facts observed
empirically is that there are only very short-range correlations
in price changes and the time memory is less than one trading day
and as small as minutes for the most liquid markets
\citep{Liu-Gopikrishnan-1999-PRE,Gopikrishnan-1999-PRE,Mantegna-Stanley-2000}.
In contrast, volatility exhibits a memory over up to the order of
one year.

If we combine the information on the time scales from the pdf's of
returns shown in Figure \ref{Fig:OC:pdfR}, the correlation of
returns shown in Figure \ref{Fig:OC:CorrR} and the correlation of
the volatility shown in Figure \ref{Fig:OC:CorrAbsR}, we can
obtain a rough idea of the correspondence between the time step of
the model and real trading time. From Figure \ref{Fig:OC:pdfR}, we
see that the pdf of returns at time scale $\tau=4$ and $\tau=16$
are similar to the empirical one at the daily scale for major
stocks and indices. This suggests that one day corresponds to
roughly $4-16$ time steps of the model. This correspondence is
compatible with Figure \ref{Fig:OC:CorrR} for the absence of
correlation at the daily time scale observed empirically. This
correspondence gives with Figure \ref{Fig:OC:CorrAbsR} an integral
time scale for the volatility correlation of about $100/4$ to
$100/16$ days, i.e., $6-25$ days. This is about a factor of ten
shorter than observed on real markets, if we believe the relevance
of the MRW and its calibration of $T$ for real markets.

\begin{figure}
\begin{center}
\includegraphics[width=14cm]{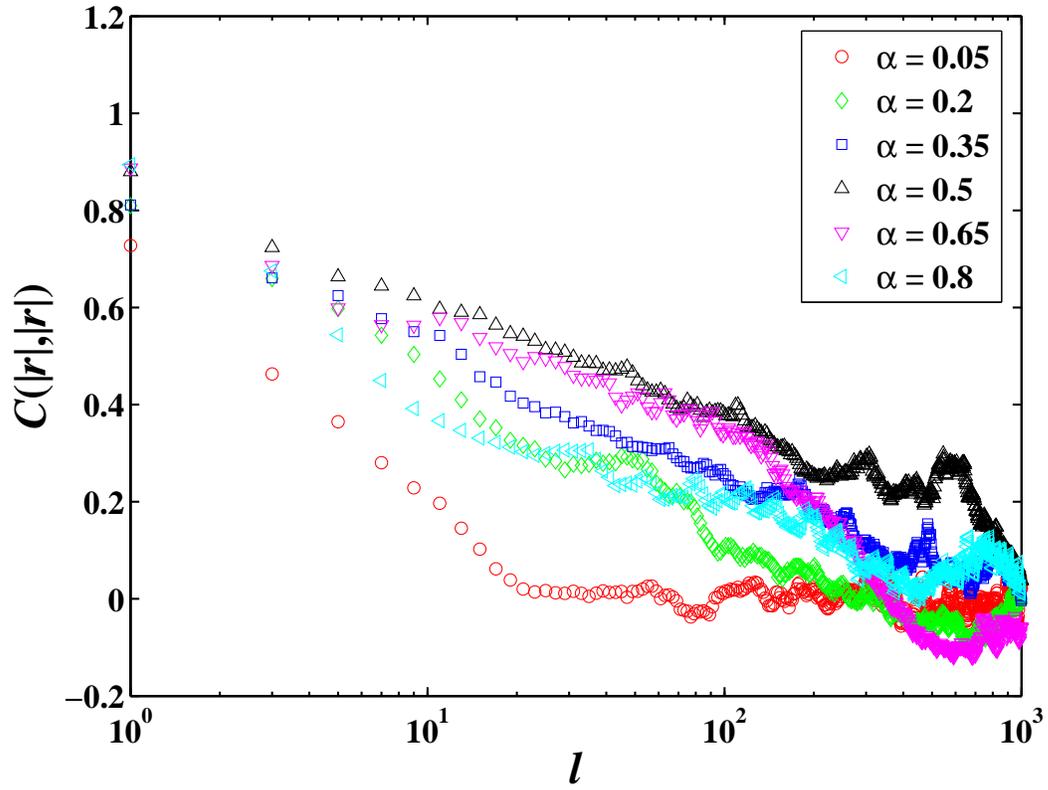}
\end{center}
\begin{center}
\includegraphics[width=14cm]{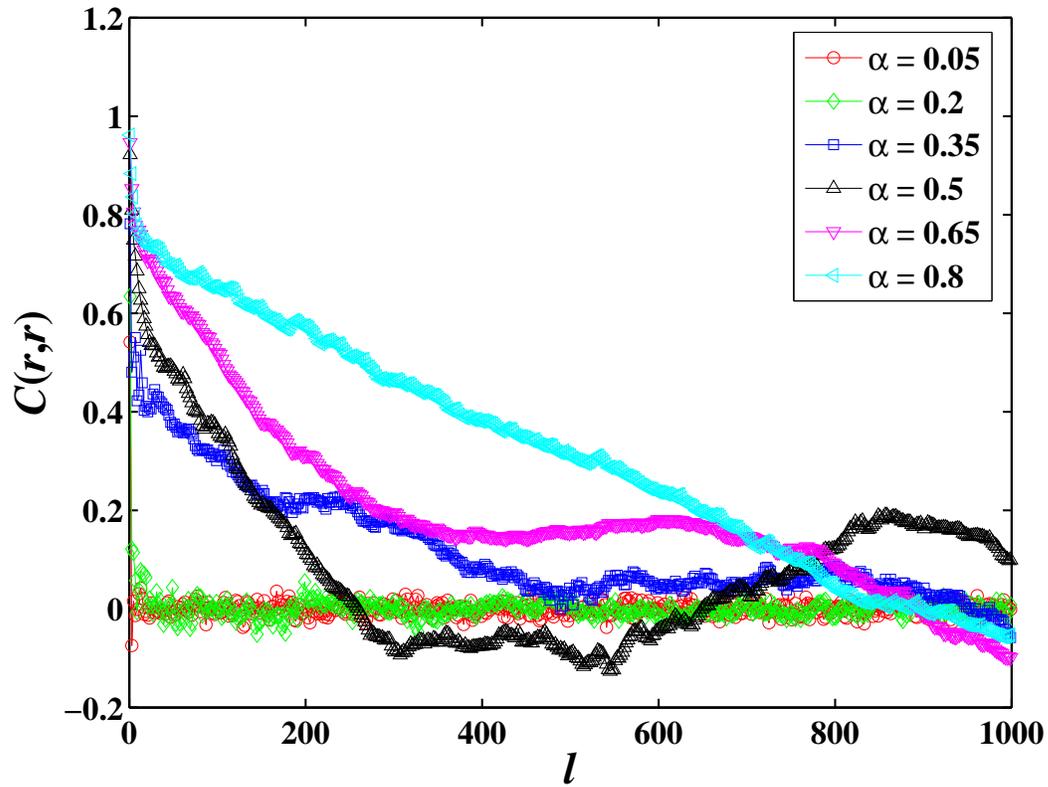}
\end{center}
\caption{The impact of $\alpha$ on the auto-correlation of the
absolute values of the returns and of the returns.}
\label{Fig:OC:alfa1}
\end{figure}

The memory of the autocorrelation of the volatility (as well as
the correlation of the returns) is sensitive to the value taken by
the parameter $\alpha$ introduced in our model, which embodies the
dependence of the coefficient $K_{ij}$ of imitation on its past
values. Figure \ref{Fig:OC:alfa1} shows that much longer ranges
for the correlation of the absolute value of returns are found for
larger values of $\alpha$, however at the cost of introducing an
unrealistic correlation of the returns. This figure suggests that
$\alpha$ cannot be larger than $0.2-0.3$ without producing
unrealistic correlation in returns.

\subsubsection{Multifractal properties}
\label{s3:MF}

\begin{figure}
\begin{center}
\includegraphics[width=14cm]{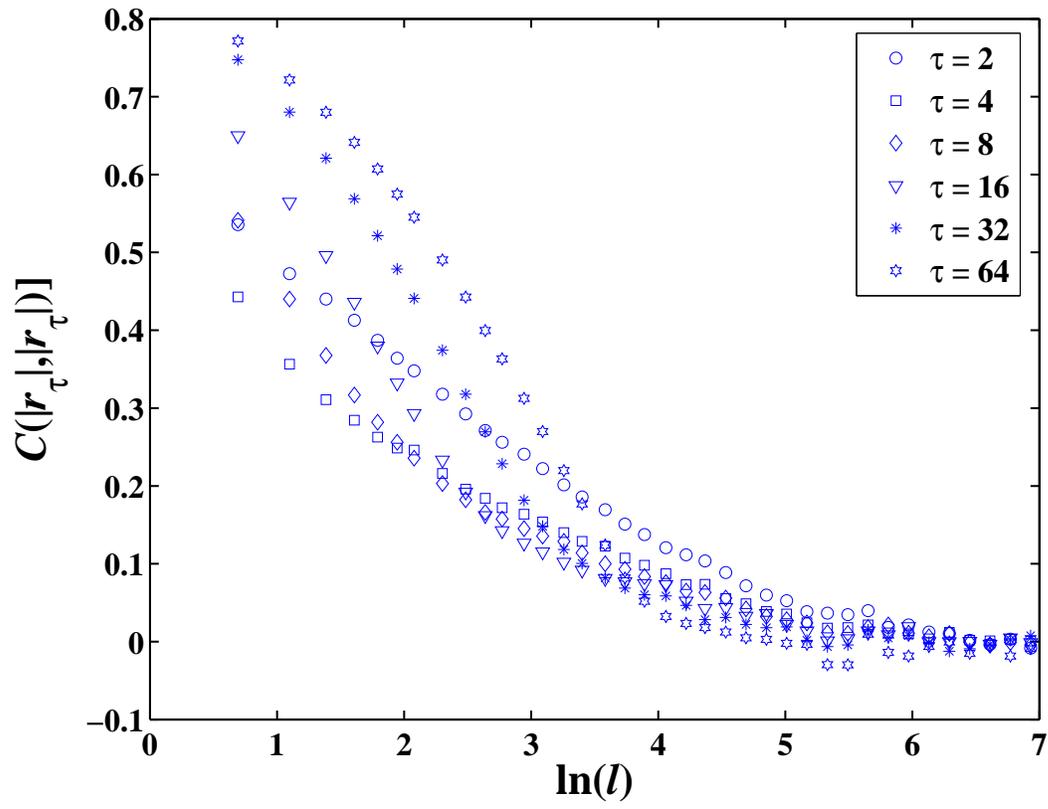}
\end{center}
\caption{Scaling of the autocorrelation functions of $|r_\tau(t)|$
for different time scales $\tau$ of the realization shown in
Fig.~\ref{Fig:OC:lnP}.} \label{Fig:OC:CRtau}
\end{figure}

The MRW also predicts (and this is well-verified by empirical
data) that the autocorrelation functions of $|r_\tau(t)|$ for
different $\tau$ should superimpose for time lags larger than
their respective $\tau$ \citep{M_etal,B_etal}. Figure
\ref{Fig:OC:CRtau} shows that this is approximately the case.

\begin{figure}
\begin{center}
\includegraphics[width=14cm]{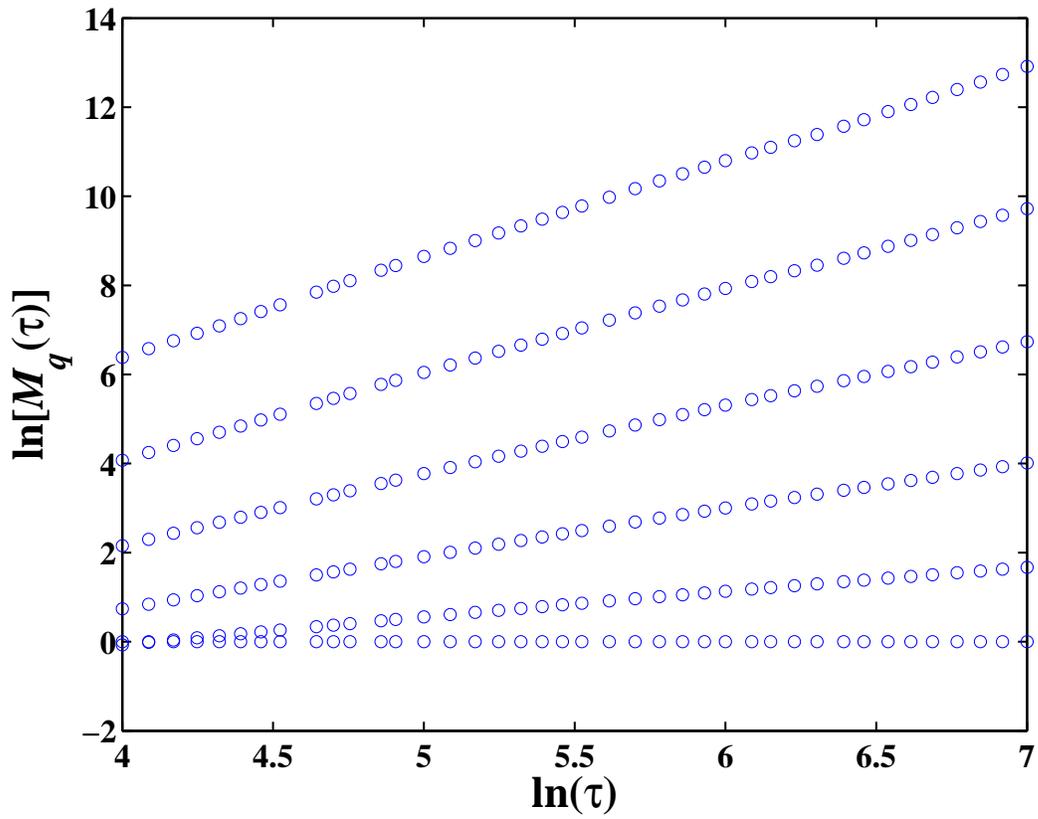}
\end{center}
\caption{Scaling of the structure functions $M_q(\tau)$ of
log-returns shown in Fig.~\ref{Fig:OC:R} at different scales
$\tau$ for different orders $q$. } \label{Fig:OC:Scaling}
\end{figure}

\begin{figure}
\begin{center}
\includegraphics[width=14cm]{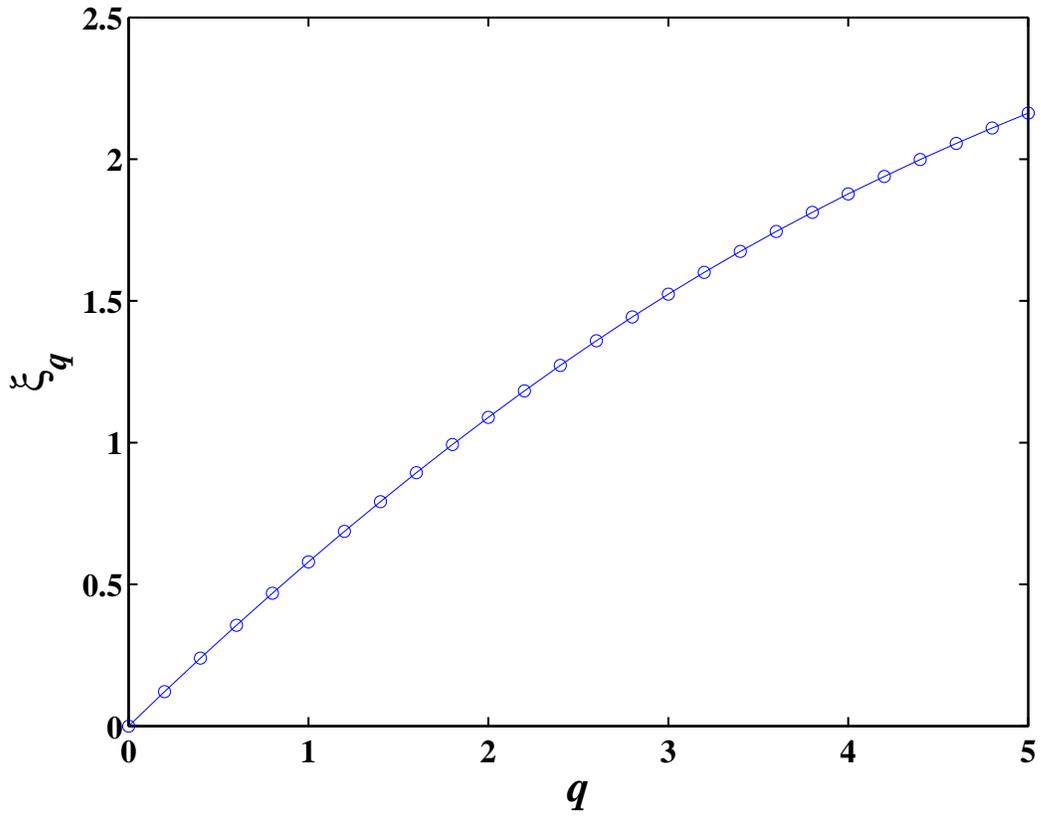}
\end{center}
\caption{Dependence of the scaling exponents $\xi_q$ defined in
(\ref{mbjowls}) as a function of the order $q$ of the structure
functions $M_q(\tau) \sim \tau^{\xi_q}$. The concavity of $\xi_q$
as a function $q$ is the hallmark of multifractality.}
\label{Fig:OC:xi}
\end{figure}

Another important stylized facts is the multifractal structure of
the absolute values of log-returns
\citep{Fisher,Mandel97,Vande1,Brachet,Bershadskii,Ivanova,Schmitt,Pasquini,MuzyQF},
which led to the proposition that the MRW might be a good model
for financial price time series
\citep{M_etal,B_etal,MuzyQF,endovol}. Figure \ref{Fig:OC:Scaling}
shows in log-log scale the structure function \be M_q(\tau) \equiv
\langle |r_{\tau}|^q \rangle \ee as a function of the time scale
$\tau$. The power law dependence of the structure functions
$M_q(\tau)$ as a function of $\tau$ is found to be reasonable. The
slopes of the lines in log-log plots give the exponents
$\xi_{\tau}$ defined by \be M_q(\tau) \sim \tau^{\xi_q}~.
\label{mbjowls} \ee Multifractality is qualified by a nonlinear
dependence of $\xi_q$ as a function of the order $q$ of the
structure function
\citep{Mantegna-Stanley-1995-Nature,Ghashghaie-1996-Nature}., as
reported in Fig.~\ref{Fig:OC:xi}.

\subsubsection{Endogenous versus exogenous shocks}

The dynamical process described by (\ref{Eq:Kimemo}) together with
(\ref{Eq:Sit}) and (\ref{Eq:rt}) describes a flux of external news
$G(t)$ which are ``digested'' by the collective behavior of the
population of traders to create a time series of returns
presenting long-range memory in the volatility and multifractal
properties, similarly to the multifractal random walk model
\citep{M_etal,B_etal,MuzyQF,endovol}. In addition, \cite{endovol}
have discovered a new consequence of multifractality in the form
of a continuous dependence of the exponent of the power law
relaxation of the volatility after a spontaneous peak as a
function of the amplitude of this peak \cite[see also other
applications in][]{endo2,endoamazon}. We proceed to test if such
an effect is found in our model.

Consider a realization giving a time series of returns $r(t)$. Let
us define the local volatility within a window $[t+1,t+\Delta{t}]$
of size $\Delta{t}$ by
\begin{equation}
 \sigma^2_{\Delta{t}}(t) = \sum_{i=1}^{\Delta{t}} |r(t+i)|^2
\end{equation}
Using this definition, we construct a time series for the
volatility $\sigma^2_{\Delta{t}}(t)$ in moving windows of size
$\Delta{t}$. The (unconditional) average volatility $E[\sigma^2]$
is nothing but the average of the time series of
$\sigma^2_{\Delta{t}}(t)$. Let us follow \cite{endovol} and
consider local burst of volatility $\sigma^2_{\Delta{t}}(t)$ with
amplitude scaled to the average volatility $E[\sigma^2]$:
\begin{equation}
\sigma^2_{\Delta{t}}(t_) = e^{2s(t)}E[\sigma^2]~. \label{mgbjle}
\end{equation}
The parameter $s$ thus quantifies the relative amplitude of a
local burst of volatility in units of the average volatility.
Following  \cite{endovol}, for a given $s$, we identify all times
$t_s$ whose volatility $\sigma^2_{\Delta{t}}(t_s)$ is close to
$e^{2s}E[\sigma^2]$, that is,
\begin{equation}
e^{2(s-ds)}E[\sigma^2] \le \sigma^2_{\Delta{t}}(t_s) \le
e^{2(s+ds)}E[\sigma^2]~,
\end{equation}
where $ds \ll 1$. For a given relative log-amplitude $s$, we
translate and superimpose all time series starting at all the
previously found times $t_s$. Averaging over these time series of
volatility obtains the average conditional relaxation function of
the volatility $E[\sigma(t|s)^2]$ following a local burst of
volatility of amplitude (\ref{mgbjle}). The multifractal random
walk model predicts a power dependence
\begin{equation}
 E[\sigma(t|s)^2]\sim t^{-\alpha(s)}~,
 \label{Eq:alpha:Def}
\end{equation}
with
\begin{equation}
 \alpha(s)=\frac{2s}{3/2+\ln(T/\Delta{t})}~,
 \label{Eq:alpha:Ana}
\end{equation}
when $\Delta{t}<t\ll \Delta{t}e^{|s|/\lambda^2}$, where
$\lambda^2\approx 0.02$. We keep the symbol $\alpha(s)$ for the
exponent in (\ref{Eq:alpha:Def}) in line with the notation of
~\cite{endovol}, but this should not be confused with the
parameter $\alpha$ in (\ref{Eq:Kimemo}) which controls the memory
of the imitation coefficients $K_{ij}$.

\begin{figure}
\begin{center}
\includegraphics[width=14cm]{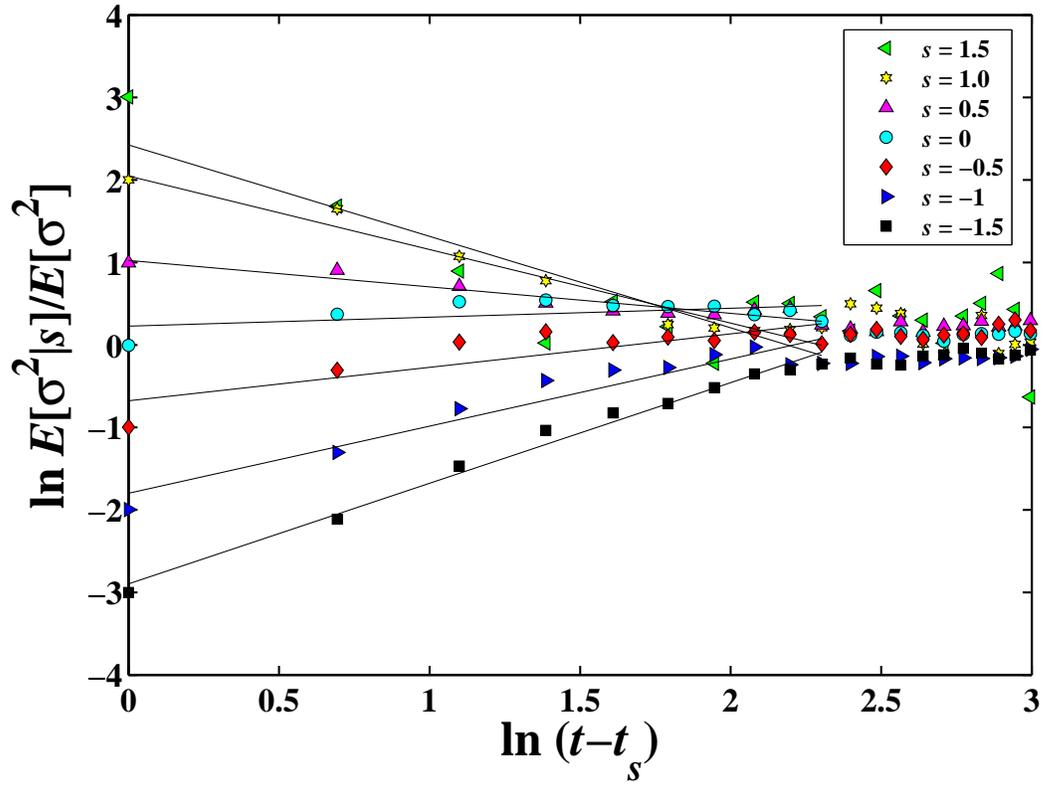}
\end{center}
\caption{Average normalized conditional volatility
$\sigma^2_{\Delta{t}}(t)/E[\sigma^2]$ as a function of the time
$t-t_s$ from the local burst of volatility at time $t_s$ for
different log-amplitudes $s$ in double logarithmic coordinates.}
\label{Fig:OC:Endo_1}
\end{figure}

\begin{figure}
\begin{center}
\includegraphics[width=14cm]{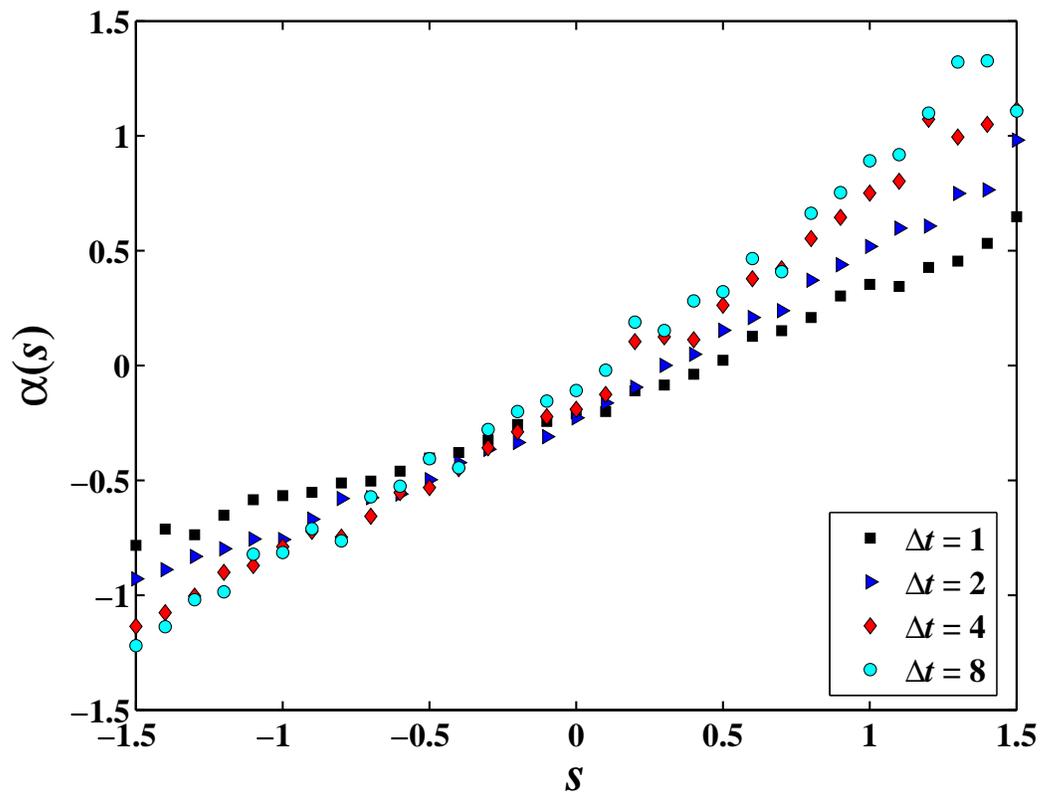}
\end{center}
\caption{Exponent $\alpha(s)$ of the conditional volatility
response  as a function of the endogenous shock amplitude $S$ for
$\Delta{t}=1, 2, 4$, and $8$.} \label{Fig:OC:Endo_2}
\end{figure}

Fig.~\ref{Fig:OC:Endo_1} shows the average normalized conditional
volatility $E[\sigma(t|s)^2]/E[\sigma^2]$ as a function of the
time $t-t_s$ to the local burst of volatility at time $t_s$ for
different log-amplitudes $s$ in double logarithmic coordinates. As
expected from the predictions (\ref{Eq:alpha:Def}) and
(\ref{Eq:alpha:Ana}) of the MRW, the average volatility after a
burst decays (respectively increases) when $s>0$ (respectively
$s<0$). Similar to real data analyzed by \cite{endovol}, we
observe approximate power laws. The power law exponents for
different values of $\Delta{t}$ are plotted in
Fig.~\ref{Fig:OC:Endo_2}. The exponent $\alpha(s)$ depends
linearly on $s$, with a slope increasing with $\Delta{t}$, in
agreement with the prediction of the MRW and the finding of
Sec.~\ref{s3:MF}.

We can actually obtain a direct estimation of the integral time
scale $T$ by studying how the slope $1/k$ of $\alpha(s)$ as a
function of $s$ depends upon $\Delta{t}$. Expression
(\ref{Eq:alpha:Ana}) predicts that $k(\Delta{t}) =
-\frac{1}{2}\ln(\Delta{t}) + \frac{1}{2}\ln(T) + \frac{3}{4}$. A
linear regression of $k$ as a function of $\ln({\Delta{t}})$ gives
$k(\Delta{t}) = -0.48\ln(\Delta{t}) + 2.08$. The first coefficient
$-0.48$ is nicely close to the exact value $-1/2$ predicted by the
multifractal theory, which provides an independent check on the
validity of multifractality. Identifying $\frac{1}{2}\ln(T) +
\frac{3}{4}$ with $2.08$ yields $T=14.3$.

\begin{figure}
\begin{center}
\includegraphics[width=14cm]{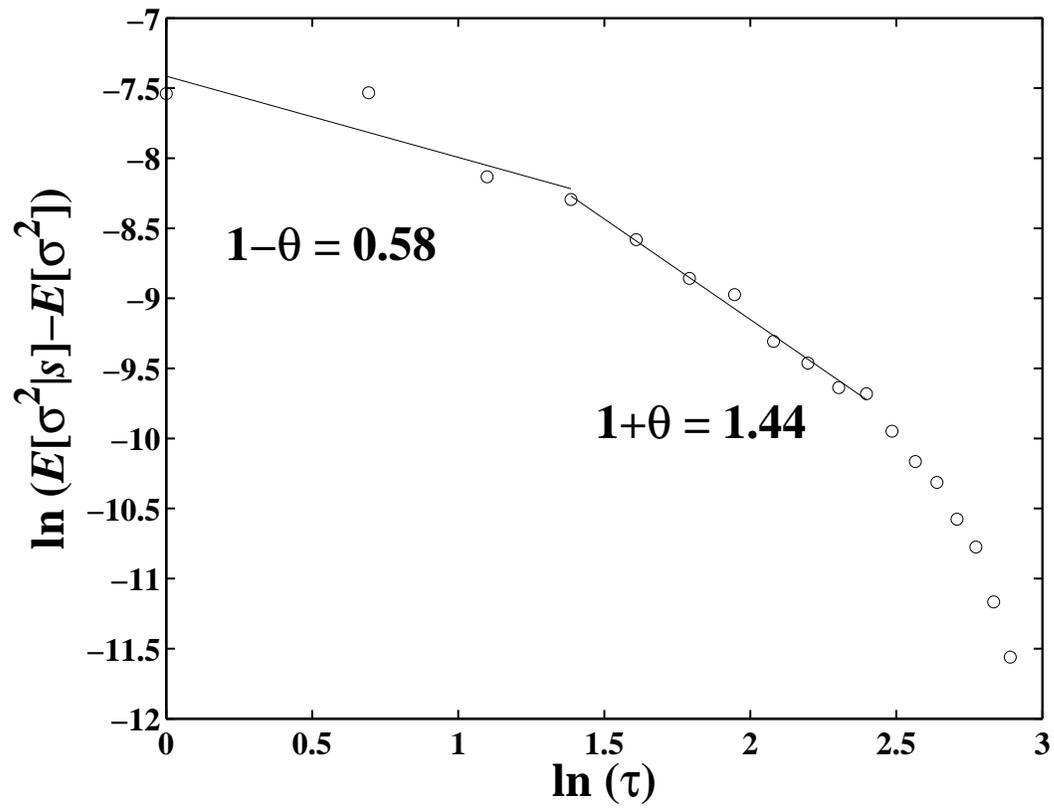}
\end{center}
\caption{Relaxation of superposed excess volatility after
exogenous shocks obtained by imposing a very large news $G(t_s)$
for $\Delta{t}=1$.} \label{Fig:OC:Exo}
\end{figure}

Fig.~\ref{Fig:OC:Exo} shows the average relaxation of the
volatility after an exogenous shock, created by imposing a very
large news impact $G(t_s)$ at a single time $t_s$ and then letting
the system evolve according to its normal dynamics thereafter. To
gather sufficient statistics, we impose such large shocks with a
periodicity of several hundred time steps, which is sufficiently
long to allow the system to relax back to its normal fluctuating
volatility. We have checked that the relaxation shown in Figure
\ref{Fig:OC:Exo} is independent of the amplitude of the shock when
sufficiently large. Note that, immediately after the news impact,
the volatility first increases over a few time step before
relaxing, which reflects the strong increase of the coupling
coefficient $K_{ij}$ and the resulting stronger cooperativity of
the agents. In order to mimic the previous analysis of the data by
\cite{endovol}, the origin of time for the relaxation of the
volatility is taken at the peak time, rather than from the
incipient exogenous news shock at $t_s$. This shifts the origin of
time by approximately $2-3$ time steps, which is compatible with
the correspondence that one trading day corresponds to $4-16$ time
steps, as discussed in section \ref{ngjs}.

According to the theory relating the endogenous relaxation to the
exogenous response function of the MRW developed by
\cite{endovol}, the relaxation shown in Fig.~\ref{Fig:OC:Exo}
should be characterized by an exponent close to $1/2$ over
approximately the same range of times $t-t_s$ as found for the
power law dependence of the relaxations shown in
Fig.~\ref{Fig:OC:Endo_1}. We indeed observe a first decay regime
which is compatible with a power law with an exponent close to
$0.5$ (but of course the range is too short to provide anything
other than an indication). We also observe a cross-over to a
faster decay, compatible with a faster decaying power with
exponent close to $1.5$. This behavior is actually expected if the
system is not exactly critical but close to critical, as shown by
\cite{endo2} and \cite{endoamazon}: the response function to an
exogenous shock should in this case cross over from a dependence
proportional to $1/t^{1-\theta}$ to $1/t^{1+\theta}$, with
$\theta=1/2$ for a multifractal system. Fig.~\ref{Fig:OC:Exo} thus
suggests that the multifractal properties of our system hold only
up to a finite time scale $T$ beyond which a cross-over to a
non-critical behavior dominates.

\section{Discussion}

In this paper, we have extended the artificial stock market model
introduced by \cite{G03AFM} to include a memory in the dynamics of
the influence coefficient on its past realization. This additional
memory turns out to be a key ingredient to reproduce the major
stylized facts of financial stock markets. In the previous
specification $\alpha=0$ of \cite{G03AFM}, the influence
coefficients $K_i$ adjust instantaneously to previous news and
returns realization. With a non-zero $\alpha$ as proposed here,
the influence coefficients exhibit an inertia. We believe that
this is a crucial property of the interactions between social
agents who only relatively slowly update their tendency to
imitate their colleagues or friends. That this parameter provides,
together with the competition between imitation and news, the main
stylized facts of financial stock markets is an encouraging sign
that we have correctly captured some of the most important
ingredients at the origin of the organization of financial stock
markets.

These ingredients comprise imitation between agents, their
influence by external news and the impact of their private
information. The imitation plus the idiosyncratic part of the
decision process give together the dynamics of the Ising model.
The news act then as an time-dependent external field. The
addition of an evolution in the influence coefficients (which can
also be called ``coupling coefficients'') make the strength of the
imitations between agents a function of the past realization of
the news and returns.

\begin{figure}
\begin{center}
\includegraphics[width=14cm]{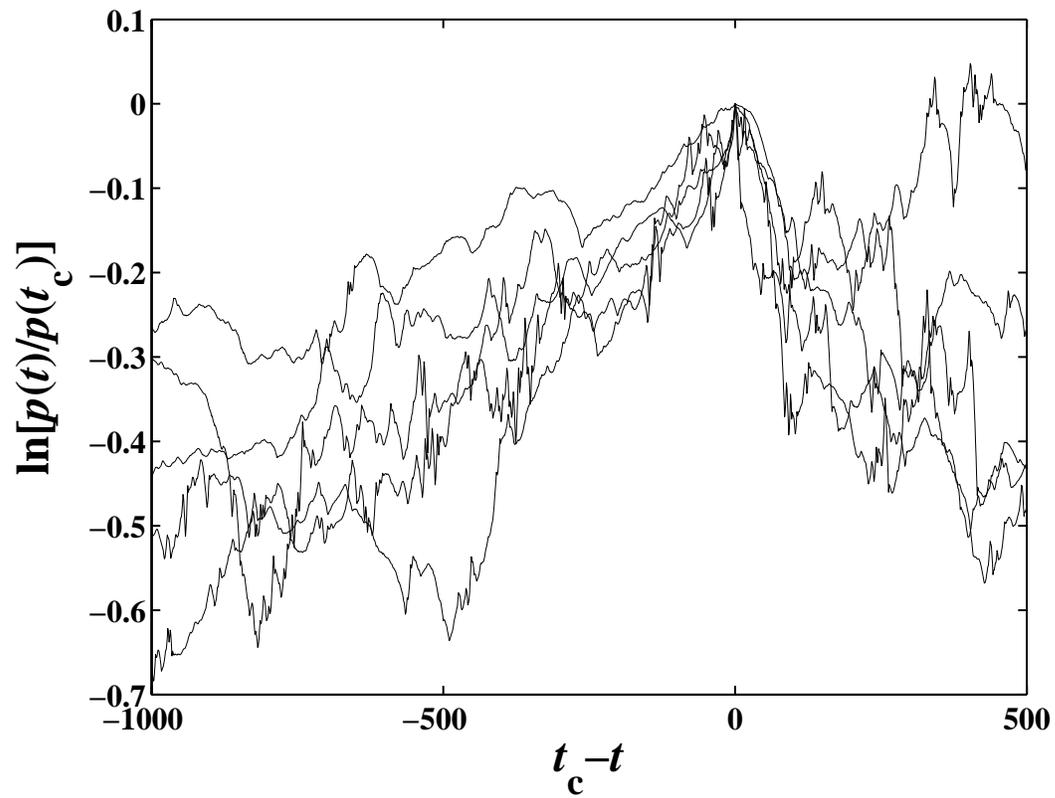}
\end{center}
\caption{Five price trajectories showing bubbles preceding crashes
that occur at the shifted time $0$. The five time series have been
translated so that the time of their crash is placed at the origin
$t=0$.} \label{FigGASM_OC_Bubbles}
\end{figure}

The empirical stylized facts of financial stock markets have been
found only for $\beta>0$ and $\alpha$ neither too small nor too
large. The condition $\beta>0$ means that agents increase the
propensity to imitate if the external news have been predictive of
the returns in the past. This behavior corresponds to agents who
misinterpret, or misattribute the source of the prediction of
returns. Alternatively, this behavior corresponds to
over-confident agents. Technically, the stylized facts in this
regime result from the fact that the model operates around the
critical point of the corresponding Ising model, with coupling
coefficients which are time-dependent and endowed with a memory of
past realizations. The critical point of the Ising model is
associated with a critical value $K_c$ for the average coupling
coefficient. Close to this value, agents organize spontaneously
within clusters of similar opinions, which become very susceptible
to small external influences, such as a change of news. This may
explain the occurrence of crashes, as argued previously
\citep[see,][Ch. 5]{Sornette-2003}. The present model exhibits
bubbles and crashes as shown in figure \ref{FigGASM_OC_Bubbles},
whose detailed study will be reported elsewhere. Intuitively, the
critical slowing down well-known to characterize the proximity to
the critical Ising point can explain the long-term memory of the
volatility while the almost absence of correlation of the returns
themselves is ensured by the impact of the news and the random
idiosyncratic decisions.

The fact that the most basic stylized facts (monomodal shape with
fat tails, short-time return correlations and long-memory of the
volatility) cannot be obtained for $\beta<0$ suggests the
importance of the imitation behavior captured by (\ref{Eq:ASMKi})
with $\beta>0$. In this model, the creation of anomalous
volatility, of its persistence, and of multifractality result from
the tendency of agents to misinterpret the combined information of
the news and of the stock market as resulting from the influence
of the agents and their imitation. Or seen from a different view
point, conditioned on their role of reflecting the stock market,
the news serve as the substrate for fostering social interactions
and reinforcing herding. By the mechanism of intermittent
reinforcing of social interaction in (\ref{Eq:ASMKi}), the
coupling coefficient $K_{ij}$ will vary and sometimes increase
close to or cross a critical value at which critical fluctuations
occur and beyond which global cooperativity dominates.

As a bonus, we have discovered that this simple model exhibits a
rich multifractal structure, diagnosed not only by the standard
convexity of the exponents of the structure functions but also by
distinct power law response functions to endogenous compared with
exogenous volatility shocks \citep{endovol,endo2,endoamazon}. To
our knowledge, this is the first nonlinear model in which such
clear distinction is documented quantitatively, based on a
bottom-up self-organization. In contrast, the multifractal random
walk which has provided the theoretical predictions used here is a
descriptive phenomenological model.

Let us end by noting the connection with the model of
\cite{Wyart}, which can be embedded in our model by putting all
$K_{ij}$ to $0$ (no imitation) and adding a dependence of
$\sigma_i$ on past correlations between the realized returns
$r(t)$  and the available information $G(t)$, similarly to the
dynamics of $K_{ij}$ in our model. \cite{Wyart} have studied the
limit of self-fulfilling conventions created by the belief of
agents on the existence of correlations between information and
returns. By this belief, traders try to estimate this correlation
from past time series and act on it, thus creating it. Our model
emphasizes the other class of conventions based on imitation and
moods. For the future, it would be interesting to combine both
mechanisms as they are arguably present together in real markets,
in order to clarify their relative importance and interplay.

\section*{Appendix: Justification that the imitation term
is rational in the absence of reliable information}

The form (\ref{Eq:Sit}) of the formulation of the decision of
agent $i$ (without the term $ \sigma_i(t) G(t)$) derives naturally
from an argument of bounded-rationality, as follows.

Let us denote $N(i)$ the number of traders directly connected to
$i$ on the graph of acquaintance ($N_i=4$ for the 2D-square
topology). The traders buy or sell one asset at price $p(t)$ which
evolves as a function of time assumed to be discrete and measured
in units of the time step $\Delta t$. In the simplest version of
the model, each agent can either buy or sell only one unit of the
asset. This is quantified by the buy state $s_i=+1$ or the sell
state $s_i=-1$. Each agent can trade at time $t-1$ at the price
$p(t-1)$ based on all previous information including that at
$t-1$. We assume that the asset price variation is determined by
the following equation \be \frac{p(t)-p(t-1)}{p(t-1)} =
F\left({\sum_{i=1}^N s_i(t-1) \over N}\right) ~+~ \sigma~
\eta(t)~.  \label{jfjfjka} \ee $\sigma$ is the price volatility
per unit time and $\eta(t)$ is a white Gaussian noise with unit
variance.  The first term in the r.h.s. of (\ref{jfjfjka}) is the
systematic price drift resulting from the possible imbalance
between buyers and sellers. The impact function $F(x)$ is such
that $F(0)=0$ and is monotonically increasing with its argument as
shown by recent empirical studies
\citep{Kempf,Pleroudemand,Lillodemand,boupotbdem}: perfect balance
between buyers and sellers does not move the price; a larger
(resp. smaller) number of buyers than sellers drive the price up
(resp. down). An often used dependence is simply a linear
relationship $F(x) = \mu x$ \citep{Beja,boucont,farmereco}. The
second stochastic term of the r.h.s. of (\ref{jfjfjka}) accounts
for noisy sources of price fluctuations. Taken alone, it would
give the usual log-normal random walk process

At time $t-1$, just when the price $p(t-1)$ has been announced,
the trader $i$ defines her strategy $s_i(t-1)$ that she will hold
from $t-1$ to $t$, thus realizing the profit/loss
$(p(t)-p(t-1))s_i(t-1)$. To define $s_i(t-1)$, the trader
calculates her expected profit $P_E$, given the past information
and her position, and then chooses $s_i(t-1)$ such that $P_E$ is
maximum. Within the rational expectation model, all traders have
full knowledge of the fundamental equation (\ref{jfjfjka}) of
their financial world. However, they cannot poll the positions
$s_j$ of all other traders which will determine the price drift
according to (\ref{jfjfjka}). The next best thing that trader $i$
can do is to poll her $N(i)$ ``neighbors'' and construct her
prediction for the price drift from this information. Note that,
in this approach, the ``neighbors'' are by definition those who
are polled by the trader according to her network of acquaintance.
The trader needs an additional information, namely the a priori
probability $P_+$ and $P_-$ for each trader to buy or sell. The
probabilities $P_+$ and $P_-$ are the only information that she
can use for all the traders who are not polled directly. From
this, she can form her expectation of the price change. The
simplest case corresponds to a neutral market where $P_+=P_-=1/2$.
The trader $i$ expects the following relative price change \be
 \mu~ \left({\sum_{j=1}^{*~N(i)} s_j(t-1) \over N}\right)~+~ \sigma~ \eta(t)~,
\ee where the index $j$ runs over the neighborhood of agent $i$.
Notice that the sum is now restricted to the $N(i)$ neighbors of
trader $i$. The contribution of all other traders, whom she cannot
poll directly, is one contribution to the stochastic term $
\sigma~ \eta(t)$. The restricted sum over the neighbors is
represented by the star symbol. Her expected profit is thus \be
\left(\mu~ \left({\sum_{j=1}^{*~N(i)} s_j(t-1) \over N}\right) +
 \sigma~ \eta(t)\right)~p(t-1)~s_i(t-1)~.
\ee The strategy that maximizes her profit is \be s_i(t-1) = {\rm
sign} \left( {\mu \over N} \sum_{j=1}^{N(i)_*} s_j(t-1) ~+~ \sigma
~ \eta(t)\right)~. \label{jgfhha} \ee The equation recovers
(\ref{Eq:Sit}) (without the term $ \sigma_i(t) G(t)$) by
identifying ${\mu \over \sigma N}$ with $K_{ij}=K$ taken uniform.
The evolution of opinions given by (\ref{jgfhha}) is nothing but
the dynamical version of the Ising model in which the sentiments
$\{s_i\}$ are called ``magnetic spins.'' Recall that the Ising
model exhibits a phase transition between two phases:
\begin{enumerate}
\item For weak coupling between spins (or large idiosyncratic
noise $\sigma$), the spins take random signs and there is no
majority opinion. The average opinion is zero. \item Above a
threshold $K_c$ for the average coupling coefficient (or below a
threshold $\sigma_c$), the spins align spontaneously along a
preferred direction; there is a non-zero majority opinion, which
can be either $+1$ or $-1$ depending on the history. The
properties of this transition and the spontaneous symmetry
breaking in the context of social agents is described in details
by \citet[][see Ch.5]{Sornette-2003}.
\end{enumerate}

In the extension (\ref{Eq:Sit22}) with (\ref{Eq:Kimemo}), the
coupling coefficients $K_{ij}$ between ``spins'' is allowed to
change with time. During their dynamical evolution, one can expect
that the coupling coefficients $K_{ij}$ between agents' sentiments
explore a large set of possible values according to a process
similar to the increments of a random walk-like trajectory. At
some times, a significant fraction of the $K_{ij}$ may approach or
pass above the critical value $K_c$: this will lead, according to
the mechanism of the phase transition in the Ising model, to the
occurrence of a majority of opinion and thus to strong herding
behavior. The introduction of the dynamics of the coupling
coefficient $K_{ij}(t)$ is reminiscent of the mechanism discussed
by \cite{SS99} in the similar context of a percolation model of
cluster of agents impacting the price evolution
\cite{Contboudaaq}, in which the connectivity parameter is varied
randomly in time. It is also providing a specific mechanism for
the occurrence of critical times as explained by \citet[][see
Ch.5]{Sornette-2003}. The additional existence of external news
corresponds in the language of the Ising model to an external
forcing ``field'' which help the opinion of the majority in the
herding phase to bifurcate towards one or the other values $\pm
1$. However, the dynamics of (\ref{Eq:Sit22}) with
(\ref{Eq:Kimemo}) is clearly more complicated than for the Ising
model due to the feedback of the external news and the collective
decision process captured by the term $\beta r(t) G(t)$ on the
coefficients of influences $K_{ij}$.

%\bibliography{MfDSI}

\clearpage
%FIGURE 1

\clearpage
%FIGURE 2

\clearpage
%FIGURE 3

\clearpage
%FIGURE 4

\clearpage
%FIGURE 5

\clearpage
%FIGURE 6

\clearpage
%FIGURE 7

\clearpage
%FIGURE 8

\clearpage
%FIGURE 9

\clearpage
%FIGURE 10

\clearpage
%FIGURE 11

\clearpage
%FIGURE 12

\clearpage
%FIGURE 13

\clearpage
%FIGURE 14

\clearpage
%FIGURE 15

\end{document}